\documentclass{article}
\usepackage[utf8]{inputenc}
\usepackage{graphicx}
\usepackage{amsmath} \usepackage{amssymb} \usepackage{todonotes}
\usepackage{stmaryrd} \usepackage{mathtools} \usepackage{textcomp} \usepackage{pgfplots} \usepackage{bm} 
\usepackage{gensymb}
\usepackage{multirow}
\usepackage{doi} \usepackage{caption}
\usepackage{subcaption}

\usepackage{authblk}

\title{Imaging and simulation-based analysis of evaporation flows over wetting edges}

\author[1]{S.~Raju\thanks{raju@mma.tu-darmstadt.de}}
\author[2]{F.~Braig}
\author[1]{M.~Fricke\thanks{fricke@mma.tu-darmstadt.de}}
\author[1]{D.~Gr\"unding}
\author[2]{E.~D\"orsam\thanks{doersam@idd.tu-darmstadt.de}}
\author[2]{H.~M. Sauer\thanks{sauer@idd.tu-darmstadt.de}}
\author[1]{D.~Bothe\thanks{bothe@mma.tu-darmstadt.de}}

\affil[1]{Mathematical Modeling and Analysis, TU Darmstadt, 64289 Darmstadt, Germany}
\affil[2]{Dep. of Printing Science and Technology, TU Darmstadt, Magdalenenstr. 2, 64289 Darmstadt, Germany}

\date{\today}

\newcommand{\dd}{{\rm d}}

\newcommand{\N}{{\nabla}}
\newcommand{\bq}{\begin{equation}}
\newcommand{\eq}{\end{equation}}
\newcommand{\ba}{\begin{eqnarray}}
\newcommand{\ea}{\end{eqnarray}}

\newcommand{\intN}{\mathbf{n}_{\Sigma}}
\newcommand{\vel}{\boldsymbol{v}}

\newcommand{\Stensor}{\mathbf{S}}

  \newcommand{\Ds}{\mathcal{D}_{\rm E}} \newcommand{\Vm}{V^{\rm (mol)}} \newcommand{\pa}{p_{\rm E}} \newcommand{\psat}{\pa^{\rm (sat)}}      \newcommand{\mobi}{{\tilde m}} \newcommand{\moleE}{n_{\rm E}^{\rm L}} \newcommand{\moleEG}{n_{\rm EG}^{\rm L}} \newcommand{\molefracE}{x_{\rm E}^{\rm L}}  \newcommand{\massfracE}{m_{\rm E}^{\rm L}} \newcommand{\massfracEG}{m_{\rm EG}^{\rm L}} \newcommand{\VliqE}{V_{\rm E}^{\rm L}} \newcommand{\VliqEG}{V_{\rm EG}^{\rm L}} \newcommand{\rhopureE}{\rho_{\rm E}^{\rm pure}} \newcommand{\rhopureEG}{\rho_{\rm EG}^{\rm pure}} \newlength\tabdist 
\newcommand{\peth}{p_{\rm E}} \newcommand{\pethsat}{p_{\rm E}^{\rm (sat)}} \newcommand{\ceth}{c_{\rm E}} \newcommand{\neth}{n_{\rm E}} \newcommand{\rhoeth}{\rho_{\rm E}} \newcommand{\xethG}{x_{\rm E}^{\rm G}} \newcommand{\xethL}{x_{\rm E}^{\rm L}} \newcommand{\Jvap}{\boldsymbol{J}_{\rm E}}   

\graphicspath{{ch4_sim/}}

\begin{document}

\maketitle

\noindent
{\bf Abstract}

\noindent We monitor the evaporation of a volatile liquid (ethanol) from an inkjet-printed liquid film, consisting of a mixture of ethanol and ethylene glycol. Interferometric video imaging technology is used for recording 2D vapor concentration profiles over the evaporating film. The vapor flow is reconstructed using numerical simulations. In this way, we reconstruct the complete flow velocity profile, and distinguish diffusive and convective gas transport flows, with quantitative tracking of the transport balances. The convective flows are driven by the buoyancy of the solvent vapor in the ambient air. In particular, we reconstruct the evaporation process from the interface of the two-component liquid. We monitor the evaporation flows, implement Raoult's and Henry's laws of vapor pressure reduction, as well as evaporation resistivity. We observe the edge-enhancement of evaporation flows at the wetting rims of the liquid film, and decompose the vapor flows in the diffusive and the convective contribution. We demonstrate how Langmuir's evaporation resistivity can be identified using vapor pressure profiles in the gas phase data and mass transfer balances.
 
\section{Introduction}

Vapor flows over planar surfaces from evaporation of liquids or of volatile components in mixtures or solutions are a key aspect of coating, printing and drying technology: Solvent-based inks, dyes or varnishes are placed on foil or sheet in a continuous or a sequence of repeating processes. The solvent fraction of gravure printing inks, typically toluene or ethanol, contributes up to 80 \% of the ink volume, and must be removed in the drying unit. Evaporation does not only imply the transition of liquid matter to the gas state. It also changes the chemical composition of the residual liquid film on the processed surface, and eventually induces secondary transport phenomena such as heat transfer, Marangoni flows inside the liquid film and the displacement of pigments as described by {\em Maki \& Kumar} \cite{Maki.2011} and in the review of {\em Craster \& Matar} \cite{Craster.2009}. Prominent is the coffee stain effect, the formation of `outline'-patterns along the edges of the liquid film, which has been studied by {\em Hu \& Larsen} \cite{Hu.2006}, or in liquid drops, as has been studied by {\em Deegan et al.} \cite{Deegan.1997}\cite{Deegan.2000}\cite{Deegan.2000b}. The tears-of-wine phenomenon on the walls of a drinking glass containing an alcoholic beverage also belongs to this class, as has been pointed out by {\em Vuilleumier et al.} \cite{Vuilleumier.1995}. In gravure or inkjet printing as it is described by {\em Kipphan} \cite{Kipphan.2000} and in the review of {\em Kumar} \cite{Kumar.2015}, evaporation of organic solvents may trigger a complex chain of instabilities, leading to fatal defects in the printed multilayer stacks such as in organic solar cells and light emitting diodes \cite{Klauk.2006}\cite{Nisato.2016}. Typically, diluted polymer solutions must be processed and dried, frequently endowed with intricate constellations of viscosity, surface tension, Marangoni numbers, solubilities, and evaporation properties, as has been pointed out by {\em Sauer et al.} \cite{Sauer.2020}. 

When liquid patterns with complex geometries are deposited on a solid surface by, e.g., inkjet or gravure printing, volatile components of the liquid film are not homogeneously evaporated over the liquid area. Rather, the evaporation flow from the edges is enhanced compared to the center. The concentration of the volatile component in the liquid film drops faster at the edges. This has multiple effects for the solidification of dissolved or dispersed material, for the wetting capabilities of the residual liquid film, and, by Marangoni effect, on its overall structural stability. On the other hand, using liquid mixtures of deliberate composition combined with appropriate conditions of the drying atmosphere, Marangoni drag could also be employed as a tool in the design of novel printing strategies as has been demonstrated by {\em Hernandez-Sosa et al.} \cite{HernandezSosa.2013}, who systematically developed solvent mixtures for polymer inks, or by {\em Davis et al.} \cite{Davis.1997}. 
From this, it is evident that a concept for continuous monitoring of the vapor flows in the drying atmosphere, and the reconstruction of diffusive as well as convective transport could be extremely useful. 

For this reason, we have developed a simulation-enhanced interferometric imaging system of vapor concentration and flow. We used inkjet printing for depositing a liquid film of a mixture of ethanol in ethylene glycol on an aluminium plate, which was then placed in the coherent beam of light of a Michelson interferometer. The ethanol vapor slightly increases the optical refraction index over that of pure air. This, in turn, shifts the phase of a coherent light beam passing the atmosphere in a way which is proportional to the molar concentration of the ethanol vapor. This causes an interference pattern which is detected by a camera chip. A 2D section of the vapor concentration profile can thus be determined optically. After some image processing steps such as phase unwrapping and a correction of optical border effects we obtained vapor concentration in the gas atmosphere over the printed sample. With these data we calculated the loss of residual ethanol from the liquid film over time, and thus obtained the initial concentration as well. By numerical simulation of this setup, and by adjusting boundary conditions with the measurement, we could reconstruct the complete vapor and air transport including buoyancy-driven convection where the mass density of pure ethanol vapor is 50 $\%$ over that of air at the same temperature and pressure. The simulations thus allow us to distinguish diffusive and convective regimes in the vapor field. With this we obtained the physical parameters which were inaccessible by the interferometric measurement itself. 

Although the plain principle is fairly simple to demonstrate, and there are deeper studies on evaporation of sessile liquid drops by {\em Dehaeck et al.} \cite{Dehaeck.2014} and by {\em Shukla \& Panigarhi} \cite{Shukla.2020}, we had to face some challenges in order to obtain really meaningful results and to match experiment and simulation in a reasonable manner. Details on the measurement system have been presented by {\em Braig et al.} \cite{Braig.2021} and \cite{Braig.PhD}. In order to obtain enhanced resolution of the vapor concentration profile at the edges of the liquid film we inkjet-printed rectangular stripes of the liquid mixture on the substrate plate such that the long edges of these stripes were oriented parallel to the propagation direction of the light beam.
The complete edge of the liquid film is thus projected to one pixel in the image obtained from the camera chip. Each 3D surfaces of equal vapor concentration is mapped to an isobaric line in the 2D projection. We also corrected the errors in the measured vapor profile which result from the ends of the rectangular films where the laser beam has to enter and leave the vapor volume. This was done by taking profiles from several such rectangles with increasing lengths so that we could correct for the `finite-size' effects. In this way we also avoided the need of an Abel transform of our image data, which would have been necessary if we had created a circular edge, as in the mentioned studies on sessile drops. Another aspect concerned the liquid formulation. While {\em Dehaeck et al.} \cite{Dehaeck.2014} considered  fluorinated liquids, and {\em Shukla \& Panigarhi} \cite{Shukla.2020} focussed on evaporating hydrocarbon drops, we used a mixture of ethanol and ethylene glycol which was matched in viscosity and surface tension for deposition by an inkjet print head. The detailed ink requirements for inkjet printability have been explained by {\em Martin et al.} \cite{Martin.2008}, and by {\em Perelaer et al.} \cite{Perelaer.2009}. This adaption was a crucial precondition for printing wetting patterns of arbitrary shapes.  

The physical origin of the enhanced evaporation from the edges of the film is essentially related to the feature that vapor transport is limited by the comparatively slow diffusion of the vapor molecules in the gas, driven by the gradient of concentration. The vapor isobars in the gas phase close to the patterned surface inherit the shape of the wetting pattern, including the sharp curvature at the edges. Isobars are close together there, and the concentration gradient exceptionally strong. Moreover, vapor pressure close to the surface reflects the local chemical potential of the volatile component at the surface. Chemical potential is not constant, but depends on the ethanol concentration in the liquid, known as Raoult's effect. At larger distance from the surface, the gradients of  vapor concentration become small, and eventually convective transport in the air takes over, induced by the buoyancy of the more dense ethanol vapor. We describe this by the Schmidt number $Sc$ and the Grashof number $Gr$, which represent the relations between the convective and the diffusive transport coefficient, and the buoyancy versus the viscous force, respectively.

\paragraph{Structure of this paper:} After introducing the hydrodynamic concept and the relevant properties of liquid mixtures in sections \ref{lab:hydrodyn} and \ref{lab:liquid}, we discuss our approach to evaporation from the thermodynamical point of view in section \ref{lab:chempot}. Also the measurement technique for vapor concentration in the gas atmosphere is explained there. The experimental setup by which we record the vapor distribution is described in section \ref{experimental}. The method of simulation and how we match this with the experimental data is shown in section \ref{simulations}. We also show here how additional information can be obtained from simulation which is not immediately accessible in the physical measurement. The results of our simulation-assisted analysis are shown in section \ref{results}. In section\ref{discussion} we give an interpretation and discuss some particular features of the method. We finally present some further options and open questions in section \ref{conclusion}.

\newpage
\clearpage

\section{Mathematical modelling}
\label{lab:hydrodyn}

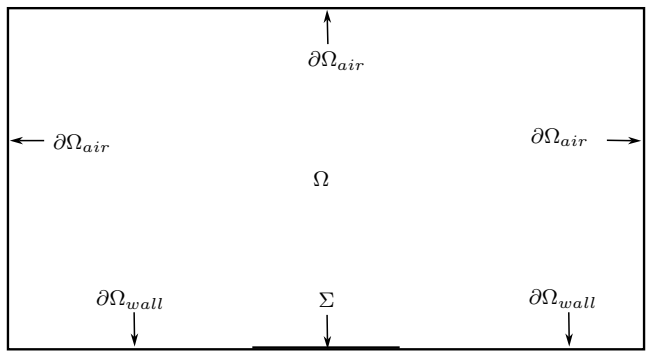
\begin{figure}[htbp]
	\centering
	\def\svgwidth{0.7\textwidth}
	{\footnotesize
		\begingroup \makeatletter \providecommand\color[2][]{\errmessage{(Inkscape) Color is used for the text in Inkscape, but the package 'color.sty' is not loaded}\renewcommand\color[2][]{}}\providecommand\transparent[1]{\errmessage{(Inkscape) Transparency is used (non-zero) for the text in Inkscape, but the package 'transparent.sty' is not loaded}\renewcommand\transparent[1]{}}\providecommand\rotatebox[2]{#2}\newcommand*\fsize{\dimexpr\f@size pt\relax}\newcommand*\lineheight[1]{\fontsize{\fsize}{#1\fsize}\selectfont}\ifx\svgwidth\undefined \setlength{\unitlength}{368.50389375bp}\ifx\svgscale\undefined \relax \else \setlength{\unitlength}{\unitlength * \real{\svgscale}}\fi \else \setlength{\unitlength}{\svgwidth}\fi \global\let\svgwidth\undefined \global\let\svgscale\undefined \makeatother \begin{picture}(1,0.5384616)\lineheight{1}\setlength\tabcolsep{0pt}\put(0.48849329,0.06777495){\color[rgb]{0,0,0}\makebox(0,0)[lt]{\lineheight{1.25}\smash{\begin{tabular}[t]{l}$\Sigma$\end{tabular}}}}\put(0.14025995,0.07144575){\color[rgb]{0,0,0}\makebox(0,0)[lt]{\lineheight{1.25}\smash{\begin{tabular}[t]{l}$\partial\Omega_{wall}$\end{tabular}}}}\put(0.81756261,0.07233968){\color[rgb]{0,0,0}\makebox(0,0)[lt]{\lineheight{1.25}\smash{\begin{tabular}[t]{l}$\partial\Omega_{wall}$\end{tabular}}}}\put(0.48011093,0.25855515){\color[rgb]{0,0,0}\makebox(0,0)[lt]{\lineheight{1.25}\smash{\begin{tabular}[t]{l}$\Omega$\end{tabular}}}}\put(0.07271019,0.31635369){\color[rgb]{0,0,0}\makebox(0,0)[lt]{\lineheight{1.25}\smash{\begin{tabular}[t]{l}$\partial\Omega_{air}$\end{tabular}}}}\put(0.8209679,0.32555497){\color[rgb]{0,0,0}\makebox(0,0)[lt]{\lineheight{1.25}\smash{\begin{tabular}[t]{l}$\partial\Omega_{air}$\end{tabular}}}}\put(0.47167751,0.44547451){\color[rgb]{0,0,0}\makebox(0,0)[lt]{\lineheight{1.25}\smash{\begin{tabular}[t]{l}$\partial\Omega_{air}$\end{tabular}}}}\put(0,0){\includegraphics[width=\unitlength,page=1]{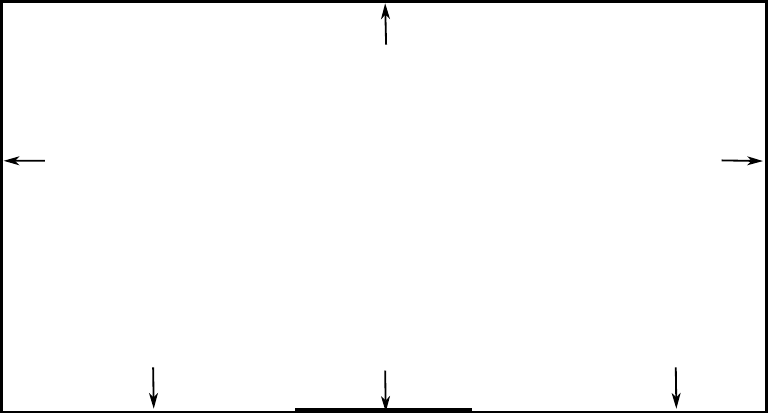}}\end{picture}\endgroup  	}
	\caption{Schematic diagram of fluid in a domain $\Omega$ with a planar interface $\Sigma$ intersecting the solid boundary of the domain $\partial\Omega_\text{wall}$ while, rest of the boundaries enclosing the fluid domain are denoted as $\partial\Omega_{air}$.}
	\label{fig:domain}
\end{figure}

We describe the evaporation of the liquid film within a domain $\Omega$ which represents the gas phase. The boundaries of $\Omega$ are represented by $\partial \Omega$ defined as $ \partial\Omega \coloneqq \{ \partial\Omega_\text{wall} \cup \partial\Omega_{air} \cup \Sigma \}$, where $\Sigma \subset \partial\Omega$ is a planar face, called the liquid film, which imposes particular boundary conditions on the flow velocity and the vapor concentration. In the remainder of $\partial\Omega$,  we impose homogeneous Neumann and Dirichlet conditions which represent the interface to the dry part of the solid wall which supports the liquid film, and the infinite gas reservoir which is assumed to be ideally mixed. The closure of the domain, $\overline{\Omega}$, is defined in \eqref{eq:fluidDom} and all the subsets of the domain are as represented in Figure \ref{fig:domain}.
\noindent
\begin{equation}
    \begin{aligned}
    \overline{\Omega} \coloneqq \{\Omega \cup \partial\Omega_{air} \cup \partial\Omega_\text{wall} \cup \Sigma  \}.  \label{eq:fluidDom}
\end{aligned}
\end{equation}
The gas in the domain $\Omega$ is a mixture of air and ethanol vapor, and characterized by its mass density $\rho$, total pressure $p$, partial pressure $\peth$ of ethanol vapor, gas flow velocity $\vel$, and the molar flux density $\Jvap$ of ethanol vapor. In contrast to an  atmosphere consisting of pure air, the transport of ethanol vapor in the mixture is partly accomplished by diffusion, and we subsume both mechanisms in the flux density $\Jvap$. If there was no diffusion, one could write $\Jvap = M_{\rm E}^{-1}\rhoeth\vel$, where the molar mass $M_{\rm E}$ of ethanol appears just for reasons of convention. Due to its finite mass density $\rho$, a vertical pressure gradient $\rho g\boldsymbol{n}_{\Sigma}$ acts on the gas, where $g$ is gravitational acceleration, and $\boldsymbol{n}_{\Sigma}$ is the normal vector on $\Sigma$. By the feature that the vapor pressure $\peth$ of ethanol is not constant in the gas domain, and because the density of ethanol vapor is enhanced by 50 \% over that of air (at the same partial pressures and temperatures), gas density $\rho$ will be slightly different from the average density $\rho_{0}$. We shall handle this using the Boussinesq approximation and assume constant total pressure $p$, temperature $T$ and mass density $\rho_{0}$. Air, as the main component of the atmosphere, is composed of several distinct partial gases. However, these are all ideal and non-condensing, in contrast to the ethanol which may pass from the liquid to the vapor state. As ethanol vapor supplants air from $\Omega$, partial pressure of air is $p-\peth$. Any particular distinction of air composition beyond the ethanol fraction is thus unnecessary. 

It is useful to equivalently describe the gas phase composition in domain $\Omega$ by the molar concentration $\ceth^{\rm G} = \neth/V = \rhoeth/M_{\rm E}$ of the ethanol vapor, where $\neth$ is the number of moles of evaporated ethanol distributed in a gas volume $V$, $\rhoeth = M_{\rm E}\peth/RT$ is the mass density of the ethanol in the evaporated state, $M_{\rm E} = 0.046\;{\rm kg}/{\rm mol}$ is its molar mass, and $R = 8.314\;{\rm J}/({\rm mol}\,{\rm K})$ the universal gas constant. Note that we take $M_{\rm E}$ in units of ${\rm kg}/{\rm mol}$ to make calculations conform with  SI standards.

The total mass, ethanol vapor mass and momentum transport equations read
\begin{equation}
\label{022_continuity}
    \partial_t \rho+\text{div}(\rho \vel)=0 \quad t>0, \boldsymbol{r} \in \Omega , \\
\end{equation}
\begin{equation}
\label{022_vaporflow}
    \partial_t \ceth^{\rm G} + \text{div} \Jvap =0 \quad t>0, \boldsymbol{r} \in \Omega , \\
\end{equation}
\begin{equation}
\label{022_NavierStokes}
    \partial_t \rho\vel +\text{div}(\rho \vel \otimes \vel)=\text{div} \Stensor+\rho g \boldsymbol{n}_{\Sigma} \quad t>0, \boldsymbol{r} \in \Omega ,
\end{equation}
where $\Stensor$ is the stress tensor related to the shear tensor $\mathbf{D}$ and to the gas viscosity $\eta$ as follows:
\begin{equation*}
    \begin{aligned}
        \Stensor = -p \mathbf{I} + \Stensor^{\text{visc}} ,
    \end{aligned}
\end{equation*}
\begin{equation*}
    \Stensor^{\text{visc}} = 2 \eta \mathbf{D}, \quad \mathbf{D} = \frac{1}{2} \left( \nabla \vel + (\nabla \vel)^{T} \right) \quad .
\end{equation*}
The representation of the liquid film by $\Sigma$ containing the condensed material needs some clarification. As the film thickness is extremely small even compared to any mesh size in the simulation, we assign all its parameters and dynamics to the interface $\Sigma$, rather than to an extra volume domain bordered by $\Sigma$. In a one-component liquid film, vapor pressure and surface tension would be the only physical parameters, and even constant under isothermal conditions. Actually, the liquid film is a mixture of ethanol and ethylene glycol. This mixture contains variable molar fractions $\xethL$ of ethanol, and $1 - \xethL$ of ethylene glycol. Saturation vapor pressure $\pethsat$ of ethanol as well as the surface tension depend on $\xethL$. 

Raoult's and Henry's laws state that the saturation vapor pressure of the volatile ethanol is reduced due to the dilution, and also surface tension is related to $\xethL$, which is the origin of a Marangoni effect in the evaporating film. In the course of evaporation, $\xethL$ and consequently $\pethsat$ will drop continuously. Proliferation of ethanol vapor terminates when the reservoir is exhausted, and the liquid film contains ethylene glycol only. This feature and its physical implications will be discussed in section \ref{lab:liquid}. 

The hydrodynamical model is completed by the boundary conditions at the domain borders. At the interface $\Sigma$, which is at rest, we assume that
\begin{equation}
   \left. \vel \right|_{\Sigma} = 0 \quad .
   \label{eq:velSigmaDirchlet}
\end{equation}
Regarding the ethanol vapor, we impose that 
\begin{equation}
\mathbf{n}_{\partial \Omega_{\rm wall}}\cdot\left.\nabla \ceth^{\rm G}\right|_{\partial \Omega_{\rm wall}} = 0
\end{equation}
on the non-wetted part of the substrate. We also need to match the boundary conditions of the ethanol concentrations $\ceth$ and $\xethL$ of the gas and the liquid phase, respectively, see eq. (\ref{evapsurfdensity}). This particular physics is discussed in section \ref{lab:chempot}.

 \subsection{Liquid film composition}
\label{lab:liquid}

The ethanol vapor in the gas phase is emanating from a liquid film of 50 $\mu$m in thickness, which is, in the present case,  a mixture of ethylene glycol and ethanol. These two liquids are miscible to any molar ratio. The surface tension depends on the molar ratio of the two components. As a consequence, a solutal Marangoni effect occurs whenever there are local gradients of concentration at the surface. We discuss this effect in detail, and show that the film represents the limiting case of a very large Marangoni number. The particular effect of concentrations on vapor pressure and evaporation is discussed later in section \ref{lab:chempot}.

The surface tension of mixtures of ethanol and ethylene glycol has been studied in \cite{Azizian.2003}. At $25^{o}C$ surface tension of this mixture decreases from $48.6\,{\rm mN}/{\rm m}$ at $\xethL = 0$ to  $22.3\,{\rm mN}/{\rm m}$ at $\xethL = 1$ in a non-linear way. The solutal Marangoni coefficient at $\xethL = 0.18$ is approximately $\sigma_{x} = -0.48\,{\rm mN}/({\rm m}\cdot\%)$ per percent of molar ethanol concentration.  The solutal Marangoni number ${\rm Ma} = \sigma_{x} b \Delta\xethL/\eta D_{\rm L}\sim 10^{6}$ of ethanol in ethylene glycol is very large, where $D_{\rm L}\sim 3\cdot 10^{-9}\,{\rm m}^2/{\rm s}$ is the diffusion coefficient of ethanol in ethylene glycol, $b \sim 3\cdot 10^{-3}$ m is the width of the liquid strip, and $\Delta \xethL$ the expected difference in molar ethanol concentration over the liquid layer width. Peclet number ${\rm Pe}_{\rm L} = b v_{\rm m}/D_{\rm L}$ of transport in the liquid phase is approximately 1500, whereas one obtains ${\rm Pe}_{\rm G} = b v_{G}/\Ds \sim 10$ in the gas phase, where $v_{G}$ is the average velocity of the gas over $\Omega$, and $\Ds \sim 1.0\cdot 10^{-5}\,{\rm m}^2/{\rm s}$ the vapor diffusion coefficient from Table \ref{tab:Gasdiffusion}. For this reason, we assume that ethanol-rich mixture with low surface tension is continuously dragged towards ethanol-poor positions with higher surface tension, and that the ethanol loss caused by evaporation is continuously replenished. The velocity of this lateral flow inside the liquid film, even in spite of the dominance of viscous friction in this film, which has a thickness of only $h \sim 5\cdot 10^{-5}\,{\rm m}$,  can be estimated as $v_{\rm m} \sim h \sigma_{x} \Delta x_{\rm E}/\eta b \sim 16\,{\rm mm}/{\rm s}$, where $\eta\sim 5\cdot 10^{-3}$ Pa\/s is the viscosity of the mixture at relevant values of $\xethL\sim 0.18$. 

\begin{table*}[ht]
	\centering
		\begin{tabular}{|c|c|c|}
			\hline
			vapor    & Diffusion coeff.  & Schmidt num. \\[\tabdist]
			              & $\Ds\;[10^{-4}\frac{{\rm m}^2}{\rm s}] $ & ${\rm Sc} = \eta/\rho \Ds$\\[\tabdist]
			\hline
			water   & $0.25^{\rm (a)}$  & $0.66$ \\[\tabdist]
			methanol & $0.14^{\rm (b)}$  & $1.14$ \\[\tabdist]
			ethanol  & $0.10^{\rm (a)}$  & $1.5$ \\[\tabdist]
			acetone   & $0.124^{\rm (a)}$ & -- \\[\tabdist]
			n-octane  & $0.05^{\rm (b)}$  & $3.2$ \\[\tabdist]
			n-decane  & $0.06^{\rm (b)}$  & $2.7$ \\[\tabdist]
			benzene   & $0.08^{\rm (b)}$  & $2.0$ \\[\tabdist]
			toluene   & $0.087^{\rm (a)}$ & -- \\[\tabdist]
			1,2-dichlorbenzene & $0.069^{\rm (a)}$ & -- \\[\tabdist]
			\hline
			${\rm O}_2$ & $0.19^{\rm (b)}$  & $0.84$ \\[\tabdist]
			He     & $0.71^{\rm (b)}$  & $0.22$ \\[\tabdist]
			${\rm H}_2$ & $0.78^{\rm (b)}$  & $0.20$ \\[\tabdist]
			\hline						
		\end{tabular}
	\caption{Gas diffusion coefficients and Schmidt numbers in air, at 23 ${}^0$C. (a): GSI Environmental Inc.\ chemical database 2010. (b): \cite{Martinez.2001}.}
	\label{tab:Gasdiffusion}
\end{table*}
The flow velocity is larger by more than one order of magnitude than the ratio of liquid strip width $b\sim 3\,{\rm mm}$ and observation time which is of order of 30 s. We therefore assume that ethanol concentration is kept almost constant over the complete width $b$ of the liquid film. Ethanol concentration is thus described by one single global value $\xethL(t)$ which depends on the age $t$ of the liquid film. Details of the experimental procedure and liquid film handling are explained in section \ref{experimental}.

\subsection{Evaporation from liquid mixtures}
\label{lab:chempot}

Material transport through the interface $\Sigma$ between the liquid and the gas-like state, as far as non-boiling situations are concerned, are usually described by the equality of chemical potentials of the volatile component. Knowing the concentration gradients in the bordering phases, this is sufficient to close the evaporation problem. However, it appears that we have witnessed another effect, known as evaporation resistance, which reduces the evaporation rates significantly below the expected value. We shall explain this below.  

Evaporation is a coexistence phenomenon of the liquid and the vapor phase of the same molecular species, i.e., of ethanol in the present case. In thermal equilibrium, and sufficiently close to it, the chemical potentials $\mu_{\rm G}$ of the vapor, and $\mu_{\rm L}$ in the liquid, are continuous at the interface $\Sigma$. If there was no transport, they ought to be constant throughout the entire domain $\Omega$ including the liquid reservoir. In the present case we impose that $\mu_{\rm G} < \mu_{\rm L}$, at least in some distance from the interface $\Sigma$. This continuously drags ethanol from the liquid bulk through the interface, either by diffusion or convection, and finally into the gas. 

Due to the dilution of the ethanol by the non-volatile ethylene glycol, $\mu_{\rm L}$ is not a constant as in pure ethanol, but a function of the molar ethanol concentration $\xethL$ in the mixture, where ethylene glycol contributes the fraction of $1 - \xethL$ to the molar liquid quantity. The concentration $\xethL$ of ethanol decreases in time $t$, and eventually approaches 0 after a while. If this is a slow process, and in particular slower than any molecular time scale, one may assume that the chemical potential $\mu_{\rm L}$ has a well-defined value at the interface $\Sigma$, and is identical to the limiting value of $\mu_{\rm G}$ at $\Sigma$. This feature may be accompanied by normal gradients of chemical potentials into both phases, which represent the diffusive flux densities. 

However, already {\em Langmuir \& Schaefer} \cite{Langmuir.1943} have recognized that this equi\-lib\-rium-rela\-ted view can lead to a substantial overestimate of evaporation rates from a liquid surface. The reason is the presence of traces of impurities which are adsorbed on the interface, and which make the interface less permeable for other molecules. This is the origin of evaporation resistivity. Forcing volatile molecules to pass the interface creates a discontinuity of the chemical potential. This discontinuity is proportional to the normal flux density through $\Sigma$. The relevance of this evaporation resistivity at liquid-gas-interfaces has been elucidated by {\em Barnes} \cite{Barnes.1986}, implicating the need to account for it in liquids with a limited purity level, or if they are contaminated by extensive contact to solid surfaces. The thermodynamic and continuum-related view on this phenomenon, which we share here, has been elaborated by {\em Bothe} \cite{Bothe.2022}. One important aim of the present study is to find a method which can distinguish this non-equilibrium phenomenon from equilibrium effects such as vapor pressure reduction or molecular affinities in the liquid phase, as well as from those non-equilibrium features which are localized at, e.g., the moving wetting line of the liquid film.

As a convention, we set the chemical potential $\mu_{\rm E}^{\rm L}$ of a liquid surface of the pure evaporating agent to 0. As this interface is in local equilibrium with the adjacent vapor phase where the agent has partial pressure  $\psat(T)$ at the given temperature $T$, one finds that the chemical potential, in units of ${\rm J}/{\rm mol}$, is
\begin{equation}
    \mu_{\rm E}^{\rm G} = RT\,\log\frac{\pa}{\psat(T)} .
\end{equation}
This could be equated with the boundary limiting value of $\mu_{\rm E}^{\rm L}$ of the chemical potential of the ethanol in the liquid film in case of equilibrium.

The chemical potential of ethanol in the liquid phase is a function of molar fractions. In a mixture of two non-interacting liquids with molar fractions $\xethL$ and $1-\xethL$, the chemical potential is reduced by $RT\log\xethL$, which is Raoult's law. Due to the enthalpy of mixture of the two components, the concentration $\xethL$ has actually to be replaced by the activity $A_{\rm E}\sim H\cdot\xethL$, where $H$ is Henry's coefficient. For the volatile component, one expects that $H>1$. For the present case of ethanol in ethylene glycol at $0 \le \xethL \le 0.3$, {\em Gil et al} \cite{Gil.2008} have obtained a substantial value of $H \approx 3$. This indicates that one should expect a pronounced enrichment of ethanol vapor in the gas phase compared to the liquid solution. This also implies a decrease of the chemical potential of ethanol to
\begin{equation}
\mu_{\rm E,Henry}^{\rm L} = RT\,\log(H\,\xethL) .
\end{equation}
Note that this is negative in the limit of small ethanol concentration. This, however, is partly counteracted by the effect of evaporation resistivity of ethanol at the liquid-gas-interface $\Sigma$, and due to a possible surfactant or impurity layer. Illustrating this from the Boltzmann perspective to molecular transport, one could consider this as an obstacle placed at the interface $\Sigma$, scattering an approaching molecule back into the liquid bulk. We describe it by an additional bulk chemical potential localized in the liquid phase, namely 
\begin{equation}
\label{evapresist}
\mu_{\rm E,s.r.}^{\rm L} \;=\; R T\,\Theta_{\Sigma}(\boldsymbol{r})\,\log \alpha_{\rm E} .
\end{equation}
$\Theta_{\Sigma}$ is the Heavyside step function which equals 1 in the liquid, and 0 in the gas phase. The parameter $\alpha_{\rm E}$, with $0 < \alpha_{\rm E} \le 1$, could be interpreted as the probability that the ethanol molecule passes the obstacle in a collision, rather than being rejected. The gradient of the additional chemical potential does not have impact on the flux densities in the bulk phases, but represents the microscopic rebounce flux of ethanol at the obstacle: ${\boldsymbol J}_{\rm E, s.r.} \sim -\nabla\mu_{\rm E, s.r.} \sim {\boldsymbol n}_{\Sigma}\,\delta_{\Sigma}(\boldsymbol{r})\,\log\,\alpha_{\rm E}$, where $\delta_{\Sigma}(\boldsymbol{r})$ denotes the delta function supported on $\Sigma$, and ${\boldsymbol n}_{\Sigma}$ is the outward normal vector on the liquid film. The total chemical potential of the liquid is
\begin{equation}
    \label{chempotliqu1}
    \mu_{\rm E}^{\rm L}(\boldsymbol{r},t) = RT\,\left(\log (H\,\xethL(t))\,+\,\Theta_{\Sigma}(\boldsymbol{r})\, \log\alpha_{\rm E}\right) .
\end{equation}
The chemical potential of ethanol in the liquid film is
\begin{equation}
    \mu_{\rm E}^{\rm L} \;=\;RT \log (H\,\alpha_{\rm E}\,\xethL) .
\end{equation}
The molar concentrations of ethanol in gas and in the liquid phases can be related according to
\begin{equation}
    \left.\xethG\right|_{\Sigma} \;=\; H\,\alpha_{\rm E}\,\left.\xethL\right|_{\Sigma} .
\end{equation}
The respective density of ethanol in the vapor phase at pressure $\pethsat(T)$ and temperature $T$ is
\begin{equation}
\label{evapsurfdensity}
    \left.\ceth^{\rm G}\right|_{\Sigma} \;=\; \frac{H\,\alpha_{\rm E}}{V_{\rm E}^{\rm (mol)}}\,\left.\xethL\right|_{\Sigma} ,
\end{equation}
where $V_{\rm E}^{\rm (mol)} = RT/\pethsat(T)$ is the molar volume and $\pethsat(T)$ is the vapor saturation pressure of pure ethanol at temperature $T$.
This is important to note, because it allows us to cross-check the values of $H$, $\alpha_{\rm E}$ and the amount of evaporated ethanol $\xethL(t)$, which is known from the initial preparation of the liquid mixture.
Concerning ethylene glycol, we neglect its evaporation because its vapor pressure is very small compared to that of ethanol, see Table \ref{tab:fluids}.

In the vapor phase, the chemical potential of ethanol vapor at partial pressure $\peth = \xethG N_{\rm Av}/\Vm RT$ is reduced accordingly. However, there is no relevant interaction between gas molecules, and there is no correction required by a Henry constant. If one considers the transport of the vapor and its progressing dilution between different points $\boldsymbol{r}\in\Omega$ as a quasi-static change of state, the chemical potential can be approximated by the space- and time-dependent potential function
\begin{equation}
    \mu_{\rm E}^{\rm G}(\boldsymbol{r},t) = RT\,\left(\log\frac{\peth (\boldsymbol{r},t)}{\pethsat(T)}\,+\,\log\xethG(t)\right) .
\end{equation}
The flux density $\boldsymbol{J}_{\rm E}$ of the vapor has contributions from convection and diffusion, and commonly has the dimension of ${\rm mol}/{\rm m}^2{\rm s}$. The diffusion part is $-\mobi_{\rm E}\N\mu_{\rm E}$, and depends on the molecular mobility $\mobi_{\rm E}$. Convective transport is proportional to the flow velocity $\boldsymbol{u}$ and molar concentration $\ceth^{G} = \xethG/\Vm$, where $\Vm = RT/p$ is the molar volume:
\ba
\nonumber
\boldsymbol{J}_{\rm E}&=&\ceth^{\rm G}\,\boldsymbol{u}\,-\,\frac{\mobi_{\rm E}}{\Vm}\,\xethG\,(1-\xethG)\,\N\,\mu_{\rm E}\\[2mm]
\nonumber
&=& \ceth^{\rm G}\,\boldsymbol{u}\,-\,\frac{\Ds}{\Vm}\,(1-\xethG)\,\N\,\xethG\\[2mm]
\label{Ev_theo_03}
&=& \ceth^{\rm G}\,\boldsymbol{u}\,-\,\Ds\,(1-\xethG)\,\N\,\ceth^{\rm G} .
\ea
We further obtain the gas diffusion coefficient $\Ds = \mobi_{\rm E}RT$ of ethanol in air, given in units of ${\rm m}^2/{\rm s}$. 

The term $\xethG (1-\xethG)$ in the diffusion flux density represents the mass action feature of diffusion as a phenomenon of stochastic molecular exchange of two distinguishable species of molecules. Far apart from boiling, $\xethG\ll 1$ is small, and $1-\xethG$ can be replaced by unity. In the particular case of ethanol vapor in an atmosphere of $p= 10^{3}\,{\rm hPa}$ at $20^{o}C$, with the saturation pressure of ethanol of $58\,{\rm hPa}$, see Table \ref{tab:vapordata}, molar fraction is always in the range $0\le \xethG \le 0.058$. 

By the continuity equation, the total vapor flow density $\boldsymbol{J}_{\rm E}$ has the property that, with $\nabla\cdot\boldsymbol{u} = 0$,
\bq
\nabla\cdot\boldsymbol{J}_{\rm E}\;=\;\boldsymbol{u}\cdot\nabla\ceth^{\rm G} - \Ds\nabla^2\ceth^{\rm G} \;=\; -\partial_{t}\ceth^{\rm G} .
\label{Ev_theo_04}
\eq
Thus we know the source strength $\nabla\cdot\boldsymbol{J}_{\rm E}$ of the {\em total} evaporation flux density.

The amount of vapor which is released through $\Sigma$ at some position $\boldsymbol{r}$ and some time $t$ into the gas domain $\Omega$ is given by the integral
\ba
 {\dot q}(\boldsymbol{r},t) &=& \boldsymbol{J}_{E}\cdot \boldsymbol{n}_{\Sigma} \quad [{\rm mol}/{\rm m}^2{\rm s}]\\[2mm]
  &=& \ceth^{\rm G}\,\boldsymbol{n}_{\Sigma}\cdot\boldsymbol{u}\,-\,\Ds\,\boldsymbol{n}_{\Sigma}\cdot\nabla\ceth^{\rm G}
\ea
The first term on the r.h.s.\ vanishes because of the boundary conditions on $\boldsymbol{n}_{\Sigma}\cdot\boldsymbol{u} = 0$ at $\Sigma$. Thus, the evaporation flux density from the liquid film at $\Sigma$ is
\bq
{\dot q}(\boldsymbol{r},t)\;=\;-\,\Ds\,\boldsymbol{n}_{\Sigma}\cdot\nabla\ceth^{\rm G} .
\eq
Thus one obtains the total loss of ethanol in the liquid film by integration across $\Sigma$, yielding
    \begin{equation*}
        \Dot{m}_{\rm E} (t) = -\,M_{\rm E}\Ds \int \displaylimits_{\Sigma} \boldsymbol{n}_{\Sigma}\cdot\nabla \ceth^{\rm G}\, \mathrm{d}o .
    \end{equation*}
    where the molar mass $M_{\rm E}$ of the vapor was multiplied to the integral in order to obtain the quantity in terms of mass units.
    Note that one does not require to know $\nabla\ceth^{\rm G}$ for closing the boundary conditions of the gas diffusion problem in $\Omega$ at $\Sigma$, because already $\left.\ceth^{\rm G}(\boldsymbol{r},t)\right|_{\Sigma}$ is known from the measurement. Rather, the gradient of concentration in the vapor enables the calculation of the loss of ethanol in the liquid film. Provided that lateral flows in the liquid film can be ignored, the local concentration of ethanol at position $x$ at time $t$ inside the liquid film ($\Theta(\boldsymbol{r}) = 1$) can be obtained from
    \begin{equation}
        \ceth^{\rm L}(\boldsymbol{r},t)\;=\;\ceth^{\rm L}(\boldsymbol{r},0)\,-\, \frac{1}{h_{\rm film}} \int \displaylimits_0^t \frac{1}{| \Sigma |} \int \displaylimits_{\Sigma} \Ds \nabla c_{\rm E}^{\rm G}\left(\boldsymbol{r},t^{\prime}\right) \cdot \intN \mathrm{d} o \mathrm{d} t^{\prime} .
        \label{eq:timeDrichlet}
    \end{equation}
    where $h_{\rm film}\sim 5\cdot 10^{-5}\,{\rm m}$ is the liquid film thickness, which we approximate by its initial value.

\subsection{Vapor concentration interferometry}
\label{lab:interferometry}
   
\begin{table}[t]
\begin{center}
    \begin{tabular}{|c|c|c|c|}
    \hline
    Gas/Vapor & Molar weight & Sat.vap.pres.& Molar refractivity  \\
         & $M_{\rm E}$ [kg/${\rm mol}]$ & $\psat$ [hPa] & $R_{\rm E}$ [$10^{-6}\,{\rm m}^3$/${\rm mol}$] \\
    \hline
    ${\rm H}_{2}$ & 0.002 & -- & $2.04^{*}$ \\
    ${\rm N}_{2}$ & 0.028 & -- & $4.484^{*}$ \\
    ${\rm O}_{2}$ & 0.032 & -- & $4.065^{*}$ \\
    ${\rm CO}_{2}$ & 0.044 & -- & $6.690$ \\
    water & 0.018 & 23.4 & 3.782\\
    methanol & 0.032 & 129 & 8.14 \\
    ethanol & 0.046 & 58 & 13.291\\
    acetone & 0.058 & 246 & 16.394\\
    ${\rm CHCl}_{3}$ & 0.119 & 213 & 21.28\\
    ${\rm CS}_{2}$ & 0.076 & 395 & 21.78\\
    \hline
    \end{tabular}
    \caption{Gas and vapor data at 20 ${}^{o}$C, 1,013 hPa. Optical data measured at $\lambda = 589\,{\rm nm}$ (Na-D line) \cite{Landolt.0402}, \cite{Gardiner.1981}. ${}^{*})$ for the binary molecule. }\label{tab:vapordata}
\end{center}
\end{table}

Our interferometric measurement of the vapor concentration $\ceth$ essentially exploits the small shifts of the refractive index $n(\ceth)$ of the vapor-air mixture. According to the Lorentz-Lorenz relation\cite{Born.1932} the index is given by
\bq
\label{Ev_theo_10}
n^{2}\,-\,1\;=\;3\,(R_{\rm E}\,-\,R_{\rm (air)})\,\ceth^{\rm G} ,
\eq
where $R_{\rm E}$ and $R_{\rm (air)}$ are the specific molar refractivities of the respective materials, defined through the molecular polarizabilities $\alpha_{\rm (mol)}$ by $R_{\rm E} = \frac{1}{3}N_{\rm A}\alpha_{\rm (mol)}$, see table \ref{tab:vapordata}. For an atmosphere at $1,001\,{\rm hPa}$, $20^{o}C$, and $50\,\%$ relative humidity, saturated with ethanol vapor at $58\,{\rm hPa}$ of partial pressure, we expected that the refractive index should be enhanced by $\Delta n_{\rm max} = 3.04\cdot 10^{-5}$ compared to pure air. The phase of a coherent wave of light with wavelength $\lambda = 532\,{\rm nm}$ which passed a distance of $2L = 120\,{\rm mm}$ in this gas would be shifted by $6.85\,\lambda$ compared to pure air. Thus, approximately seven interference fringes should separate image regions of maximal vapor concentration from those with pure air. This fits very well with our instrument calibration  in section \ref{experimental}, see the P1-curve in figure \ref{exp2}, where we indeed observe seven fringes in the $L =60\,{\rm mm}$ arrangement. A 12-bit digitalization of the camera signal with 4096 luminance levels should yield a vapor pressure resolution not worse than $1.5\,{\rm Pa}$ of ethanol. In particular the phase shift from the vapor-air mixture is 
\bq
\label{Ev_theo_11}
\Phi(x,y,t)\;=\;\frac{3\,(R_{\rm E}-R_{\rm (air)})}{2\lambda}\,\int_{0}^{2L}\,\ceth^{\rm G}(\{x,y,z\},t)\,\dd z ,
\eq
where $L$ is the distance of the light beam to pass in the vapor, and $\boldsymbol{r}=\{x,y,z\}$. Note that in our experiment the beam must pass the vapor twice. We assume that $L\gg\lambda$. When this wave is superimposed with the parallel reference wave, the $z$-integrated concentration $\ceth^{\rm G}(\boldsymbol{r},t)$ at position $\boldsymbol{\xi} = (x,y)$ of the cross section of the beam of light is mapped to the camera chip. The light intensity of the interference pattern is
\bq
\label{Ev_theo_12}
I(\boldsymbol{\xi})\;=\;I_{\rm (bg)}(\boldsymbol{\xi})\,+\,I_{\rm (if)}(\boldsymbol{\xi})\,\cos\left(\Phi(\boldsymbol{\xi})\,-\,\boldsymbol{\kappa}\cdot\boldsymbol{\xi}\right) ,
\eq
$\boldsymbol{\kappa} = (\kappa_{x},\kappa_{y})$ are the vertical and horizontal tilting angles of the image plane versus the wave front. Tilting is used to create a periodic carrier pattern of parallel stripes with wave vector $\boldsymbol{\kappa}$, which is phase-modulated by the physical signal $\Phi(\boldsymbol{\xi})$ as is needed for further image processing. The absolute intensities $I_{\rm (bg)}$ and $I_{\rm (if)}$ of the background and interference pattern were determined in the empty vapor chamber. In order to extract $\Phi(\boldsymbol{\xi})$ from $I(\boldsymbol{\xi})$, a phase unwrapping algorithm has to be applied to the recorded interference pattern, restoring the absolute phase from the $2\pi$-periodic interference pattern. Details of this algorithm are described in \cite{Braig.2021}.

The properties of a selection of gases and vapors, as well as on their molecular refraction, are given in Table \ref{tab:vapordata}.

\newpage
\section{Experimental procedure}
\label{experimental}

\begin{figure}[htbp]
    \centering
    \includegraphics[width=9cm]{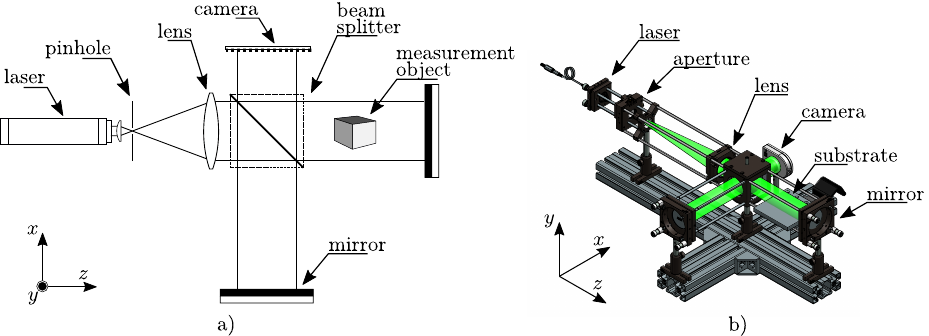}
    \caption{Left: principal Michelson interferometer setup with laser, beam splitter, CMOS camera chip, mirrors and object chamber. Right: CAD model. From {\em Braig}. \cite{Braig.PhD}}.
    \label{expsetup}
\end{figure}

Our vapor concentration measurement was based on a Michelson interferometer for coherent light phase shift measurement in the vapor atmosphere. The setup is depicted in Figure \ref{expsetup}. We restrict on the principal features of the setup, and refer to {\em Braig et al.} \cite{Braig.2021} and in particlular to \cite{Braig.PhD}, where design features of the experiment, image data processing, calibration, and the extraction of the physical quasi-2D vapor concentration profile in the evaporation chamber are described in detail. 
The laser beam from the Thorlabs CPS532-C2 laser diode (light wavelength $\lambda = 532\,{\rm nm}$, light intensity $P = 0.9\,{\rm mW}$ was expanded from 3.5 mm to a collimated beam of 42 mm in diameter. The beam was split into two orthogonal secondary beams of equal intensities, using a cubic beam splitter. One of these beams passed the evaporation zone, whereas the other one served as the reference beam. The superimposed beams were recorded by a monochromatic laser diagnosis 4.2 megapixel CMOS camera (Beamage 4M from Gentec Electro Optics Inc., Canada; sensor size 11.3 mm, pixel resolution 5.5 $\mu$m, maximum frame rate 6.2 fps, buffer memory for 64 images), with a neutral density filter N0.5 from the same manufacturer, but without any further lenses in the optical path. The in\-ter\-fero\-me\-ter was assembled using the standard cage system from Edmond Optics, and mounted on the bottom plate of the robot cell as described below. In order to keep air convection small we placed the whole equipment under a hermetically closed glass cube of approximately $1\,{\rm  m}$ in size, and to delegated mechanical operation to an industrial robot arm manipulator. This also assured  a reproducible transport of the evaporation source, and avoided long-lasting air turbulences which could have been critical for our particular experiment. Substrate positioning tolerance was 20 $\mu$m, i.e.\ substrate surface position was reproducible on the camera image within a distance of four camera pixels.   

The evaporation samples, and in particular the thin liquid films from which the vapor emanates into the gas phase, were prepared using a Xaar inkjet print head (Xaar 1003 GS6 from Xaar Plc, U.K., 1,000 nozzles, nozzle density 360 per inch, drop volume 6 to 42 pL). The print head was driven using a head interface board HIB-XR-1002/3 from Global Inkjet Systems, U.K.), which was controlled by a print managing board PMB-C2 of the same manufacturer. Fluid supply was achieved with a membrane pump PML 11458-NF 60 with BLDC option from KNF Neuberger GmbH, Germany, and a particle filter SCF-3112-J100 from Pall Corp. US, which assured a continuous fluid supply free of possible air contaminations. The membrane pump was controlled using two pressure sensors from SSI Series, First Sensor AG, Germany, and a micro controler Teensy 3.6 (PJRC, US). This proved to be useful in order to maintain a constant ink pressure in the printhead nozzles and reliable drop ejection over long time. Print head programming was done using print server and Atlas software provided by Global Inkjet Systems. 

\begin{table}[ht]
	\centering
		\begin{tabular}{|ccc|c|c|}
			\hline
			Property  & & & ethanol & ethylene glycol   \\[\tabdist]
			\hline
			 density & ${\rm g}/{\rm cm}^3$  & & 0.78 & 1.11 \\[\tabdist]
			 molar mass & ${\rm g}/{\rm mol}$ & & 46 & 62  \\[\tabdist]
			 molar volume & ${\rm cm}^3/{\rm mol}$ & & 59.0 & 55.9    \\[\tabdist]
			 sat. vap. pressure & Pa & 20${}^{o}$C  & 5,800 &  7.0  \\[\tabdist]
			                         && 27${}^{o}$C  & 8,895 &  17.0 \\[\tabdist]
			\hline
		\end{tabular}
	\caption{General properties of the used fluids, at 20${}^{o}$C}
	\label{tab:fluids}
\end{table}

As an evaporation agent adequate for inkjet printing we prepared a mixture of ethanol ($30\,{\rm wt.}$-\% or $x_{\rm A} = 0.366$) and ethylene glycol ($70\,{\rm wt.}$-\%) (both technical purity, from Sigma Aldrich), see Table \ref{tab:fluids}. Evaporation could be entirely assigned to ethanol, whereas ethylene glycol is practically non-volatile, see the vapor pressures in table \ref{tab:fluids}. Due to the dilution of the ethanol in the liquid, also the vapor concentration $\ceth^{\rm G}$ of ethanol was smaller than that over a pure ethanol film, and decreased even further as the molar concentration $\xethL$ continuously decreased in time. For this reason it was not possible to assume that the ethanol vapor pressure at the beginning of the in\-ter\-fero\-met\-ric measurement was still equal to that of the freshly printed film. For control, we thus calculated the actual ethanol vapor pressure and flux density at the liquid interface from the interferometric data. Assuming Raoult's law of vapor pressure reduction, we estimated that the ethanol concentration at the start of the measurement was roughly at $24\,{\rm wt.}$-\%  or $x_{\rm A} = 0.278$ of ethanol. This appeared to be reasonable regarding of the very low thickness of the liquid film, the printing time and the transport times of the sample from the print head to the interferometer. 

The evaporation substrate was made from a massive quadratic aluminium plate in order to prevent a temperature shift by evaporation enthalpy. Edge length was 11 cm and thickness was 1 cm. The substrate was carefully cleaned and degreased by an isopropanol treatment. Starting the experiment, a sequential set of linear stripes of the fluid was deposited by the print head, with 3 mm in width and 25 to 70 mm in length was deposited. Lateral pixel resolution was 360 dpi, and a printing velocity 40 ${\rm mm}/{\rm s}$. Twenty cycles of deposition where subsequently performed. The volume of the drops ejected from the nozzles was 12 pL, roughly. This yielded a liquid layer thicknesses of nominally 50 $\mu$m, which required 55 s of printing time, and disregarding a possible loss of ethanol by evaporation. Temperature and ambient pressure inside the experiment container were 27.1$\pm$0.5~${}^{o}$C and 1,001 hPa, respectively, and air humidity was 55 to 58 $\%$. The prepared substrate was then moved to the observation position in the interferometer beam such that the long edges of the rectangular stripes were aligned parallel to the propagation direction of the beam. Interference intensity was then recorded for 25 s, with a frame rate of 2.5 images per second. The tilt angle of the mirror in the reference beam of the in\-ter\-fero\-me\-ter was adjusted such that a parallel carrier fringe pattern with wave number $\kappa_{x} = 9\,{\rm mm}^{-1}$, see eq. (\ref{Ev_theo_12}), appeared in the camera. By demodulating the region-of-interest in Fourier space, typically $11 \times 11$ mm in size, subsequent phase unwrapping and edge correction, we obtained a continuous phase profile $\Phi(\boldsymbol{\xi})$ from each frame. Such phase profiles were recorded from sequences of evaporation stripes of different lengths $L$, and were superimposed and compared with each other, pixel by pixel. From this, the increase of phase shift per unit length of the printed stripes was determined. In this way we could distinguish and remove the contribution of the vapor emanating from the ends of the stripes, as this could have distorted the 2D projection of the vapor profile. We thus could reconstruct the signal contribution from the center section of the stripes, and calculate $\ceth^{\rm G}(\boldsymbol{\xi},t)$ from eq.~(\ref{Ev_theo_11}), as the $z$-integral there was now trivial.

\begin{figure}[htbp]
    \centering
    \includegraphics[width=9cm]{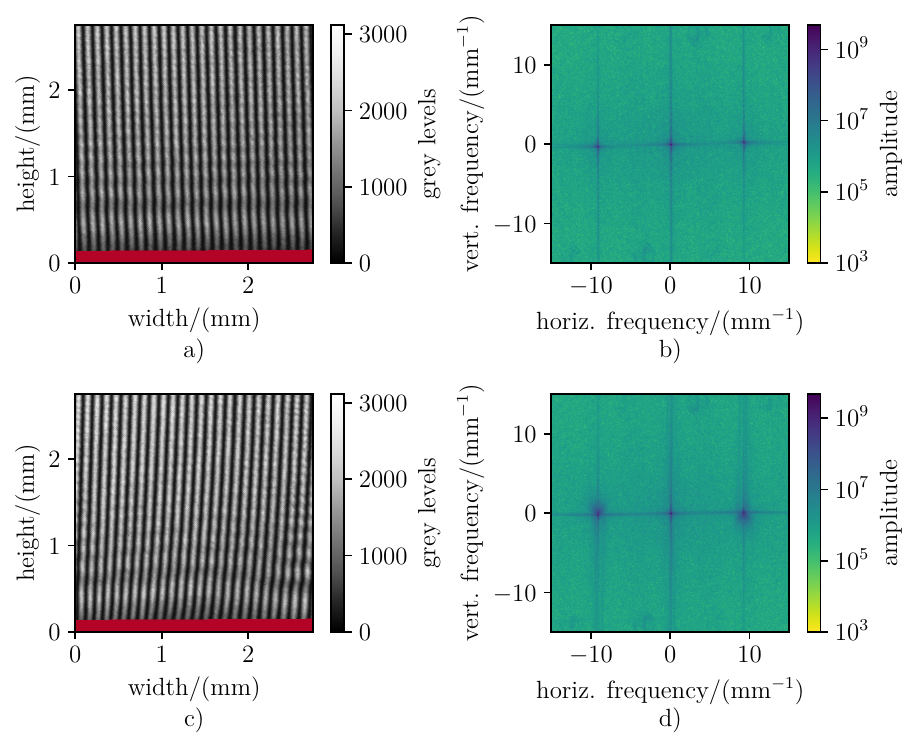}
    \caption{Optical interference patterns showing the phase-modulated carrier wave. (a): without vapor, (c) with some vapor in the bottom region. (b) and (d) show the respective Fourier representations. From {\em Braig} \cite{Braig.PhD}}.
    \label{exp1}
\end{figure}

Figure \ref{exp1}a shows the recorded interference pattern in a section ($3 \times 3$ mm) of the raw camera image ($11\times 11$ mm). The image was recorded in absence of any vapor. The interference fringes can be controlled in distance and orientation by means of the tilt angle $\boldsymbol{\kappa}$. Images showing a well-defined vertical sequence of interference patterns were preferred  for algorithmic reasons of image processing. Figure \ref{exp1}b displays the Fourier-transformed image, where the lines appear as discrete points in wave vector space. Figure \ref{exp1}c was recorded in presence of some ethanol vapor, showing a line pattern distorted by the phase shifts in the vapor atmosphere. Correspondingly, the Fourier image Figure \ref{exp1}d shows a broadening of the previous peaks. The information used for our analysis is entirely contained in the intensity distribution in the vicinity of the intensity points in the reference image, and possible error and edge refraction effects can be corrected by subtracting the phase information. 

\begin{figure}[htbp]
    \centering
    \includegraphics[width=9cm]{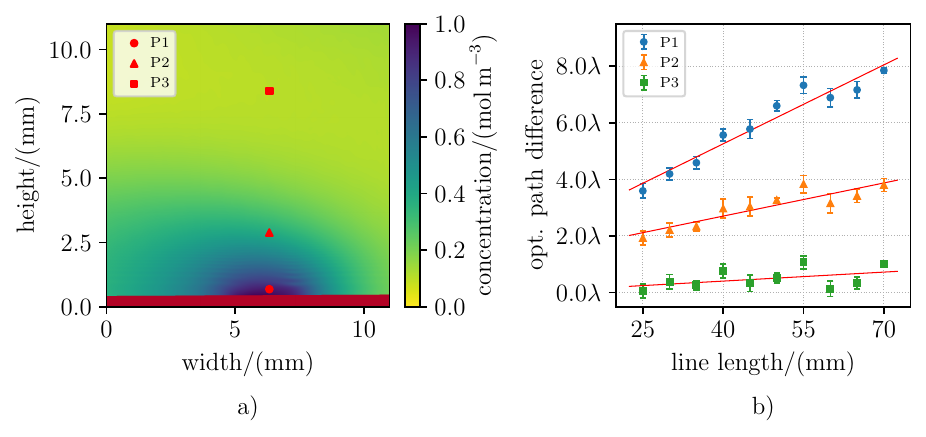}
    \caption{Vapor concentration cross section over a printed stripe (a). (b) shows the total optical phase shift (after phase unwrapping) at three different points of the profile, as functions of the lengths of the wetted stripes. From {\em Braig}. \cite{Braig.PhD}}.
    \label{exp2}
\end{figure}

Figure \ref{exp2}a shows the quasi-2D vapor distribution over the substrate, with all imaging corrections applied. The three red dots in this image represent some representative positions directly above, in short and in larger distance of the liquid surface, where a sequence of phase shifts were recorded over rectangular stripes of lengths between 25 and 70 mm. These concentrations are plotted in Figure \ref{exp2}b. As expected, the phase shift increases linearly with the length of the fields. However, the extrapolation of the fitting lines to zero length exhibits a finite offset. This offset was interpreted as the contribution of the end sections of the liquid layers to evaporation, and subtracted at any pixel from our concentration profiles.

\newpage
\section{Simulations}
\label{simulations}

The computational setup of the problem is accomplished within the open source finite volume method based framework OpenFOAM (git tag OpenFOAM-v2206) \cite{OpenFOAMv2206}. We have modified a pre-existing solver within this framework called \textit{buoyantBoussinesqPimpleFoam}. A constitutive equation for the buoyancy term in the momentum equation is introduced in terms of the passive scalar, describing the ethanol concentration, so that buoyancy-driven flow due to changes in the concentration field can be accounted for. An overview of the governing equations within the computational domain along with the boundary conditions are given below. \\
\\
For an incompressible isothermal fluid in the domain $\Omega$, the mass balance equation has the form 
\begin{equation}
    \nabla \cdot \vel = 0 , \end{equation}
where $\vel$ ($\rm m$) is the fluid velocity. The momentum balance is written as 
\begin{equation}
    \frac{\partial \vel}{\partial t}+\nabla \cdot(\vel \otimes \vel)=-\frac{1}{\rho_{0}}(\nabla p-\rho \boldsymbol{g})+\nabla \cdot\left( \nu \left(\nabla \vel+(\nabla \vel)^{\sf T}\right)\right) ,
\end{equation}
where $\rho$ $\rm (Kg/m^3)$ is the density of the fluid, $\rho_{0}$ is the density of pure air and $\nu = \eta/\rho_{0}$ ($\rm m^2/s$) is the kinematic viscosity of the fluid. The balance equation for ethanol within the gas phase read as
\begin{equation}
\frac{\partial \ceth^{\rm G}}{\partial t} + \N\cdot (\ceth^{\rm G}\,\vel) -  \,\Ds\,\N\,\ceth^{\rm G} = 0,
\label{eq:advec-diff}
\end{equation}
where $\Ds$ ($\rm m^2/s$) is the diffusivity of ethanol inside the vapor phase A and $\ceth^{\rm G}$ ($\rm mol/m^3$) is the corresponding concentration of the vapor phase. The buoyancy term $\left(\frac{\rho}{\rho_{0}}\right)$ in the momentum conservation equation is assumed to vary linearly with the concentration and is written as
\begin{equation}
 \frac{\rho(\ceth^{\rm G})}{\rho_{0}}=1 + \frac{(M_{\text{E}} - M_{\text{air}} )\ceth^{\rm G}}{\rho_{0}},
 \label{eq:concConstiLaw}
\end{equation}
where $M_{\text{E}}$ and $M_{\text{air}} $ are the molar masses of ethanol and pure air. The derivation of \eqref{eq:concConstiLaw} is given in Appendix \ref{app:constBuoyancyTerm}. \\
\noindent
A mix of homogeneous and non-homogeneous boundary conditions are used on the boundaries of the domain $\Omega$ defined in \eqref{eq:fluidDom}. The velocity at $ \partial\Omega \coloneqq \{ \partial\Omega_\text{wall} \cup \partial\Omega_{air} \cup \Sigma \}$ is determined by the no-slip condition
\begin{equation}
    \vel\big |_{\partial\Omega} = 0 .
\end{equation}

\noindent
The concentration at the boundaries of the domain depends on the type of boundary and reads as
\begin{equation}
    \begin{aligned}
        \ceth^{\rm G}\big |_{\Sigma} = \frac{H_{\rm mod}}{V_{\rm E}^{\rm (mol)}}\,\xethL ,\ \
        \ceth^{\rm G}\big |_{\partial\Omega_{air}} = 0,  \ \
        \frac{\partial \ceth^{\rm G}}{\partial y}\bigg |_{\partial\Omega_\text{wall}} = 0 ,
    \end{aligned}
    \label{eq:concBoundaries}
\end{equation}
where we have defined a modified Henry constant $H_{\rm mod} = \alpha_{\rm E} H$ taking account of the evaporation resistivity parameter $\alpha_{\rm E}$.
The new name is given by us for the solver incorporating all the above equations and boundary conditions, is henceforth called \textit{speciesBoussinesqPimpleFoam}.

Since the experimental measurements in this work are done on a 2D plane, a pseudo-2D computational setup is created to validate those measurements. The computational domain is the same as in Figure \ref{fig:domain} and has a unit cell thickness in the direction perpendicular to the 2D plane.
The length of the interface $\Sigma$ is 3 mm and the total size of the simulation domain is 203 $\times$ 100 $\text{mm}^2$. The experimental vapor field at the beginning of the measurement is used as the initial vapor field at time zero of the simulation. As the experimental field is available only in a region of $11 \times 11~\text{mm}^2$, a linear extrapolation is applied to extend the concentration field to the entire computational domain. To be consistent with the boundary condition, we choose the linear extrapolation such that the concentration vanishes at the outer boundary $\partial\Omega_{air}$. The initial experimental vapor field is asymmetrical and, hence, even though the domain is symmetric about the $y$ axis at the centre of the interface boundary, the simulation is done on the entire domain. \\
\noindent
At the end of every time step, the concentration at the interface $\Sigma$ is estimated from the loss of moles of ethanol due to the diffusion of ethanol mass out of $\Sigma$. The new reduced concentration is updated on $\Sigma$ and used as initial concentration for the next time step. The reduced concentration at the end of every time step is calculated as
\begin{equation}
    \ceth^{\rm G}(t) = \frac{\psat}{RT} H_{\rm mod} \molefracE(t).
    \label{eq:concGasPhase}
\end{equation}
The calculations are detailed in Appendix \ref{append:reducedConcCalc}. The initial concentration at the interface $\Sigma$, $ \ceth^{\rm G}(0)$ is calculated from the available experimental data and the $H_{\rm mod}$ has to be determined. This is further explained with the results in section \ref{sec:simResults}.

\section{Results}
\label{results}

\subsection{Experimental}
\label{expresults}

In this section, we present the principal results of the measured vapor concentrations in the evaporation process, the data which we needed to define the boundary conditions for the simulation, and some additional experiments we made to probe our model. First, we consider a sequence of vapor concentration profile in a typical run of the experiment. The quasi-2D corrections as described in section \ref{experimental}, see fig.~\ref{exp2}, have already been applied.

\begin{figure}[htbp]
    \centering
    \includegraphics[width=9cm]{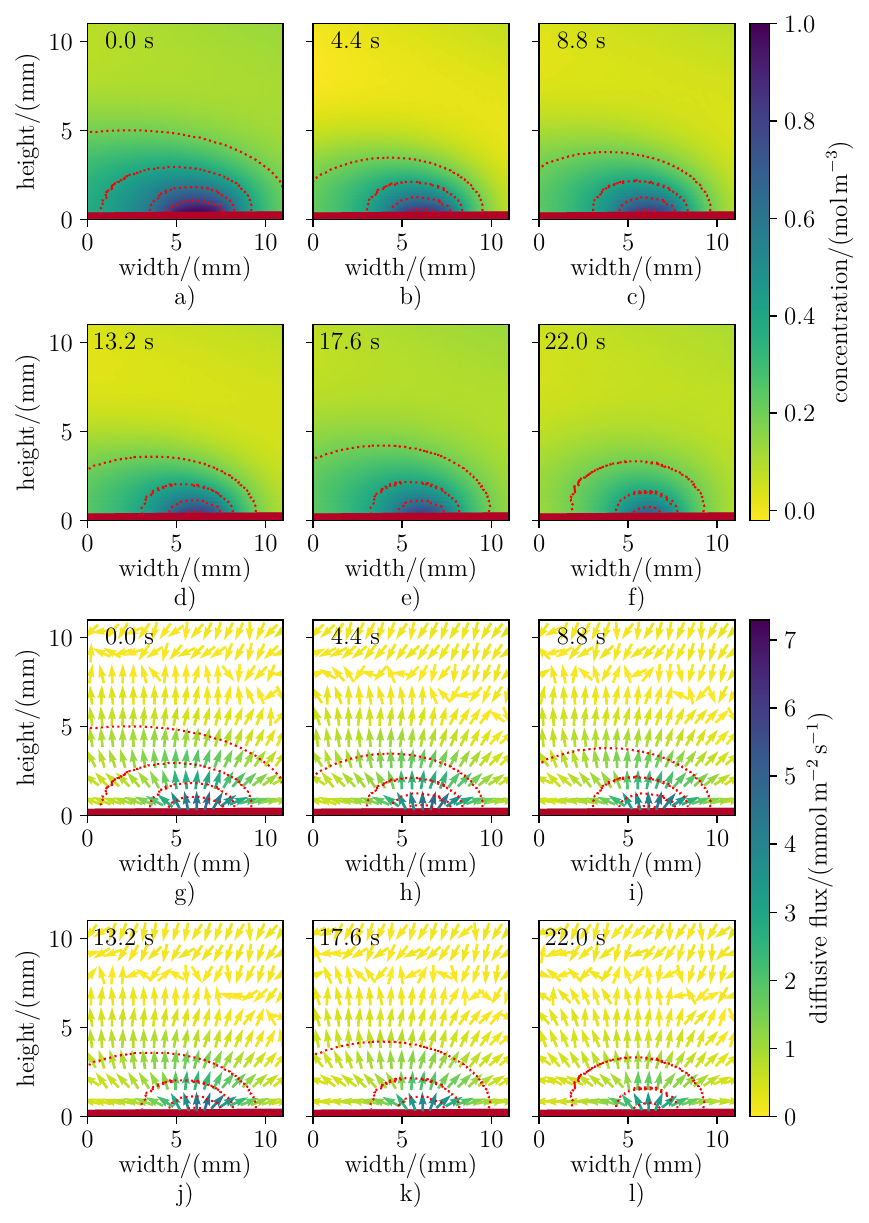}
    \caption{A sequence of six vapor concentration profiles $\ceth^{G}(\boldsymbol{\xi},t)$ over the liquid film, taken over a time span from $t = 0$ to $t = 22$ s (a--f). The red dotted lines are vapor isobars at 0.8 (inner), 0.6, 0.4, and $0.2\;{\rm mol}/{\rm m}^3$ (outer). Figures (g--l) display the normal vectors on the isobars. From {\em Braig} \cite{Braig.PhD}}.
    \label{exp3}
\end{figure}

Figure \ref{exp3}a--f shows the vapor concentration over a time interval of 22 s, with vapor isobars added in a mutual distance of $0.2\;{\rm mol}/{\rm m}^3$. Vapor concentration $\ceth^{\rm G}$ is of order of $0.9\;{\rm mol}/{\rm m}^3$ close to the liquid surface, and drops with increasing distance. With ongoing evaporation the isobars shrink, and vapor production decreases until the ethanol in the liquid film is exhausted. Outside of the wetted area, the isobars terminate at the dry surface, and their tangents intersect it in normal direction. Diffusive vapor flow density is parallel with the substrate surface here, whereas it is perpendicular over the liquid interface. The corresponding normal vectors on the isobars are shown in the corresponding Figures \ref{exp3}g--l. This feature, of course, implies that the vector field of the vapor flow density must have a singularity at the borders between wetted and dry surface.
\begin{figure}[htbp]
    \centering
    \includegraphics[width=9cm]{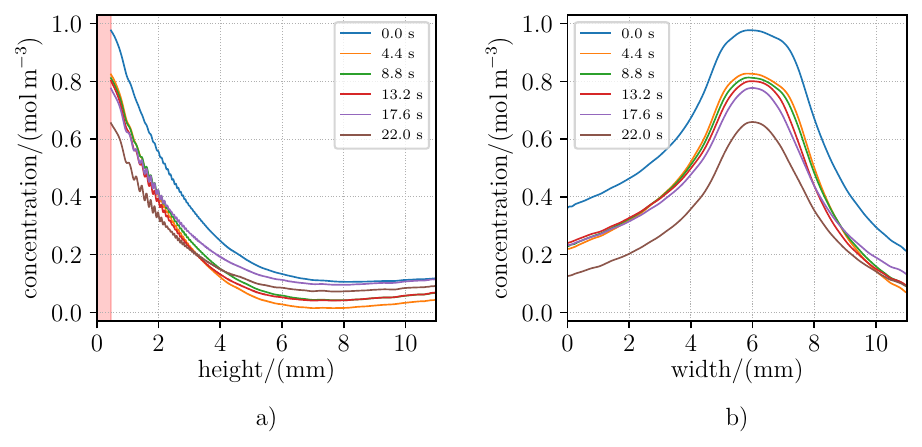}
    \caption{Ethanol vapor concentration $\ceth^{\rm G}$ as a function of distance $y$ over the substrate center (a), and across the substrate ($x$-direction) (b), at different time steps of the evaporation process. The liquid film is located between the positions $4.5\,{\rm mm}$ and $7.5\,{\rm mm}$. From {\em Braig} \cite{Braig.PhD}}.
    \label{fig:VaporProfiles}
\end{figure}

Fig. \ref{fig:VaporProfiles}, Part a), shows the time evolution of vapor concentration profile over the center of the liquid film. Vapor pressure drops rapidly with distance $y$, and approaches an almost constant value for $y > 6\,{\rm mm}$. Again, we observed a general decrease of vapor pressure in time. The observation that the vapor pressure is constant for $y > 6\,{\rm mm}$ indicates that vapor transport could not be driven by diffusion anymore, but it appears plausible that buoyancy-driven convection became more important here than diffusion.

Part b) of this plot displays the time evolution of the ethanol vapor concentration $\xethG(x_{\rm c},y,t)$ over the center of the substrate at $x_{\rm c} = 6\,{\rm mm}$, and the profile $\xethG(x,0,t)$ immediately over the substrate plane. The borders of the liquid film are located at $x = 4.5\,{\rm mm}$ and $x = 7.5\,{\rm mm}$. Vapor concentration $\ceth^{\rm G}$ shows a clear maximum of almost $ 0.99\,{\rm mol}/{\rm m}^3$ for $t = 0$ at the center of the liquid film ($x = 6\,{\rm mm}$). This corresponds to a partial pressure of $2.47\,{\rm hPa}$. 

The vapor concentration measurement close to the interface, in combination with the total amount of evaporated ethanol, leads us to the suspicion of an evaporation resistance phenomenon. This is for the following reason. For a liquid film of pure ethanol we would have expected a concentration of $3.5\,{\rm mol}/{\rm m}^3$ here, or $8.895\,{\rm hPa}$. However, we found $2.47\,{\rm hPa}$ only. One could attribute this to vapor pressure reduction by the dilution of the ethanol. Using Raoult's law and assuming a concentration of 0.366 mol-\% of ethanol in the liquid film, one should have expected $1.28\,{\rm mol}/{\rm m}^3$ in the gas phase. However, Raoult's law is inapplicable for the mixture of ethanol and ethylene glycol. As we know from \cite{Gil.2008}, Raoult's relation needs to be corrected by a considerable Henry coefficient of order of $3.0$ for this particular mixture. This would imply that the vapor concentration is $3.84\,{\rm mol}/{\rm m}^3$ at the given ethanol concentration in the liquid film. This value still disregards two effects: the loss of ethanol from the liquid film prior to the beginning of the interferometric measurement, and a possible evaporation resistance due to an adsorption layer on the liquid-air interface. The initial loss of ethanol could be recognized by integration of the evaporation flux over the measurement time. It was thus possible to distinguish the two suspected effects from our data. 

This measurement of the total evaporation flux was accomplished by integration of the normal gradient of vapor concentration $\partial_{y}\ceth^{\rm G}(x,y,t)$ in a close proximity of $y = 0.11\,{\rm mm}$ to the interface $\Sigma$. We assumed that the vapor flow close to the surface was diffusive, with neglegible convective contribution. As a reference calculation, consider a stripe of fresh mixture, with $60\times 3\times 0.05\;{\rm mm}^3$ in size. We found that 1.36 mg of ethanol have evaporated from this film drying the observed duration of the film. On the other hand, we know that 2.74 mg of ethanol where deposited there by the print head. This ratio gives us the evaporative ethanol loss prior to the start of the measurement.  

We concluded that half of the ethanol had already disappeared before the interferometric measurement started. This was to be expected regarding the length of the printing process. Repeating the above calculation with the actual ethanol quantity available, we would have had an initial ethanol vapor concentration of $1.93\,{\rm mol}/{\rm m}^3$, contrasting to the $0.99\,{\rm mol}/{\rm m}^3$ which we actually observed. This inevitable implies that the chemical potential of the ethanol on the vapor side of the interface was significantly lower than that in the liquid phase. 

We can also exclude a temperature drop across the interface as a possible reason. Temperature drop was estimated not to exceed $- 1.0 {}^{o}$C within measurement time. This would explain not more than 5 \% in relative vapor pressure drop. For this reason we claim that an evaporation resistance on the liquid interface was present as explained in the modeling section. Our data becomes consistent if we assume a reduced interface penetration probability of about $\alpha_{\rm E}\approx 0.6$ for an ethanol molecule colliding there, see eq. (\ref{evapresist}). A possible origin of this effect will be discussed in the final section.

\begin{figure}[thb]
    \centering
    \includegraphics[width=6cm]{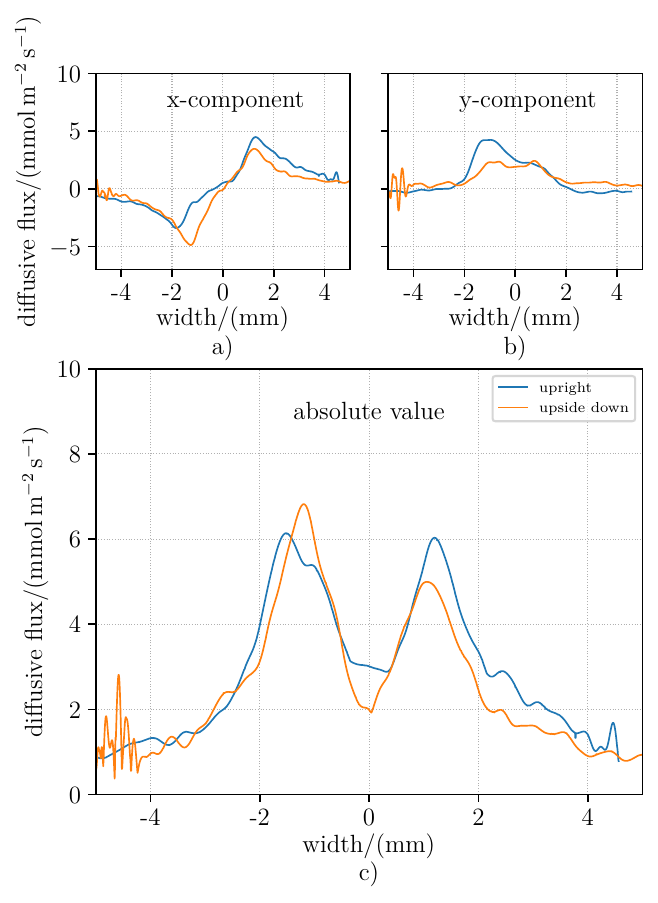}
    \caption{Ethanol evaporation flux density for upright and upside down orientations of the substrate in the interferometer. a): vapor flux density parallel to the substrate, in a distance of $0.1\,{\rm mm}$; b) flow density normal to the substrate; c) absolute value of the flux density across the substrate. From {\em Braig} \cite{Braig.PhD}}.
    \label{fig:evapangles}
\end{figure}

We shall now analyse the vapor concentration profile in lateral direction over the substrate. One observes a maximum value at the center of the liquid film, whereas the vapor concentration drops to 80 \% of that value at the rims. This indicates that ethanol concentration in the liquid film was not constant as well. This could be a consequence of the excess of evaporation expected at the boundaries. This excess of evaporation was observed in both experiment and in simulation, see figures \ref{fig:evapangles} and \ref{fig:diffFluxinYdir}. A further possible explanation is that the complete Marangoni-driven replenishing of ethanol across the liquid film cannot be fully effective. Outside the boundaries of the liquid film, vapor pressure drops rapidly with increasing distance. At the end of the recording period at $t = 22\,{\rm s}$ the vapor pressure drops quite rapidly, indicating that almost all the ethanol is gone.
Fig. \ref{fig:evapangles} shows the diffusive flux close to the substrate. Part a) is the in-plane component close to the substrate, b) the normal flow component, and c) the absolute value, which gives an approximation of the total evaporation rate per substrate area. This is based on the assumption that the vapor transport close to the substrate is purely diffusive, and that convection could be neglected.

In order to probe this assumption, which is essential for our interpretation, we repeated the experiment with the substrate face oriented downwards. This should not have affected vapor diffusion, but an inverted direction of gravity should add a substantial driving force to the buoyancy-driven convection in the gas phase. 

Fig. \ref{fig:evapangles}, part c), shows the flow density profiles for the two substrate orientations. The profiles are quite comparable in shape. We measured a total evaporation rate of $9.3\,\mu{\rm mol}/{\rm m}^{2}{\rm s}$ for the normal substrate position, and only $8.8\,\mu{\rm mol}/{\rm m}^{2}{\rm s}$ for the inverted case which is even slightly smaller. Although the discrepancy is close to the tolerances of the measurement, we agree that future clarification on the detailed structure of the diffusion zone over the substrate might bring new insight here.

\newpage
\clearpage
\subsection{Simulation results}
\label{sec:simResults}
\begin{figure}[htbp]
    \centering
    \includegraphics[width=0.75\linewidth]{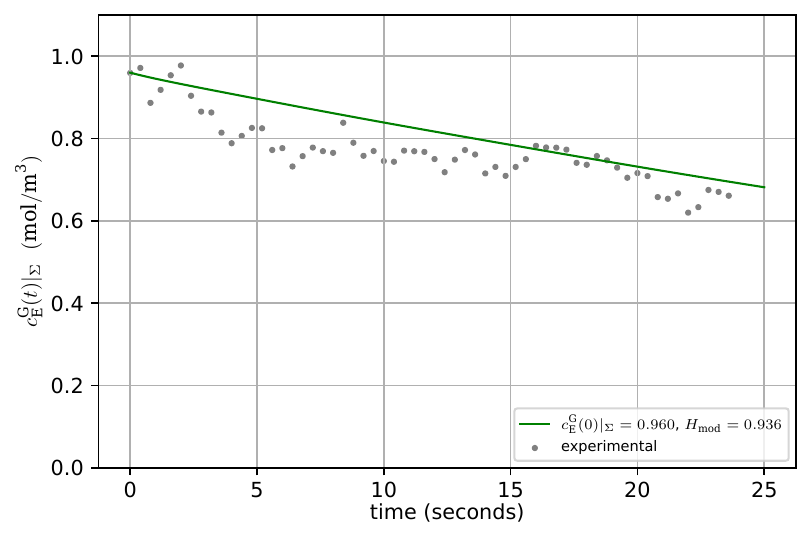}
    \caption{Evolution of concentration with an initial value of $\ceth^{\rm G}(0) = 0.960\;{\rm mol}/{\rm m}^3$ at the interface. This value is found to be the best fit starting from the mean concentration of the first experimental measurement.}
    \label{fig:conc_0.960C}
\end{figure}

\begin{figure}[htbp]
    \centering
    \includegraphics[width=0.75\linewidth]{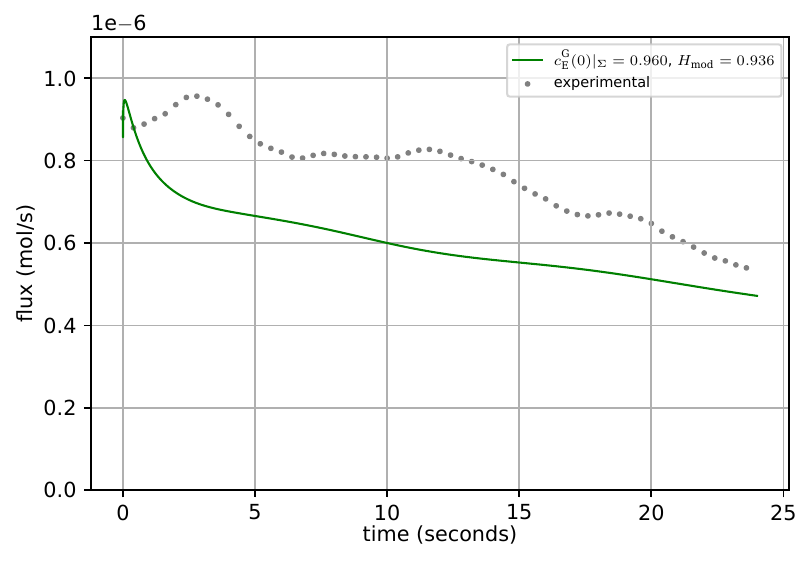}
    \caption{Evolution of diffusive flux with an initial value of concentration $\ceth^{\rm G}(0) = 0.960 \;{\rm mol}/{\rm m}^3$ at the interface. }
    \label{fig:flux_0.960C}
\end{figure}

\begin{figure}[htbp]
    \centering
    \includegraphics[width=0.75\linewidth]{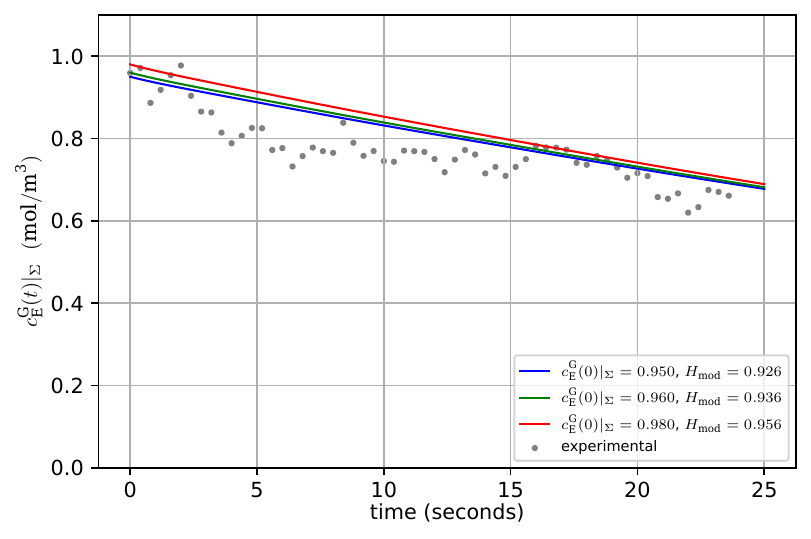}
    \caption{Evolution of concentration at the interface in the gas phase with time. Different initial concentrations around the first experimental value at time $t=0$ are chosen for which the modified Henry coefficient is calculated according to equation \eqref{eq:concGasPhase}. }
    \label{fig:concEvolution}
\end{figure}

\begin{figure}[htbp]
    \centering
    \includegraphics[width=0.75\linewidth]{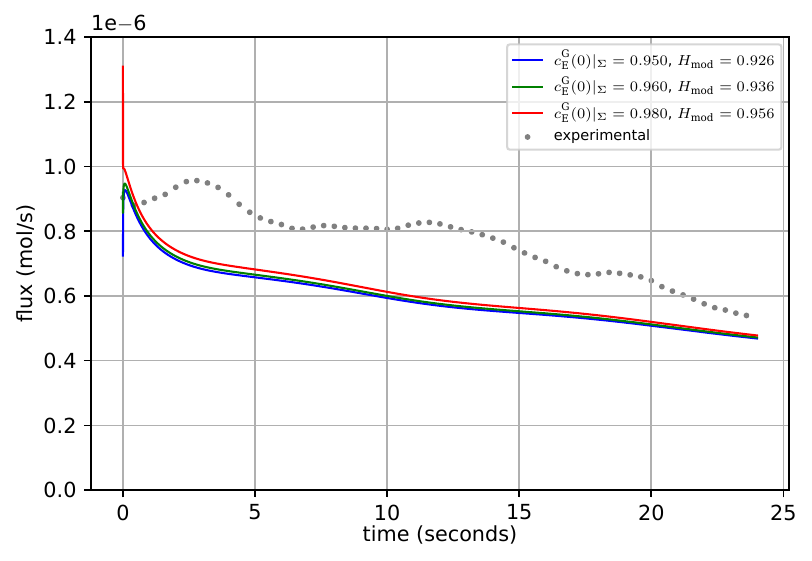}
    \caption{Evolution of diffusive fluxes out of the discretised interface compared with the fluxes evaluated from experimental measurements.}
\label{fig:fluxEvolution}
\end{figure}

\begin{figure}
\begin{subfigure}[htbp]{0.45\linewidth}
\includegraphics[width=\linewidth]{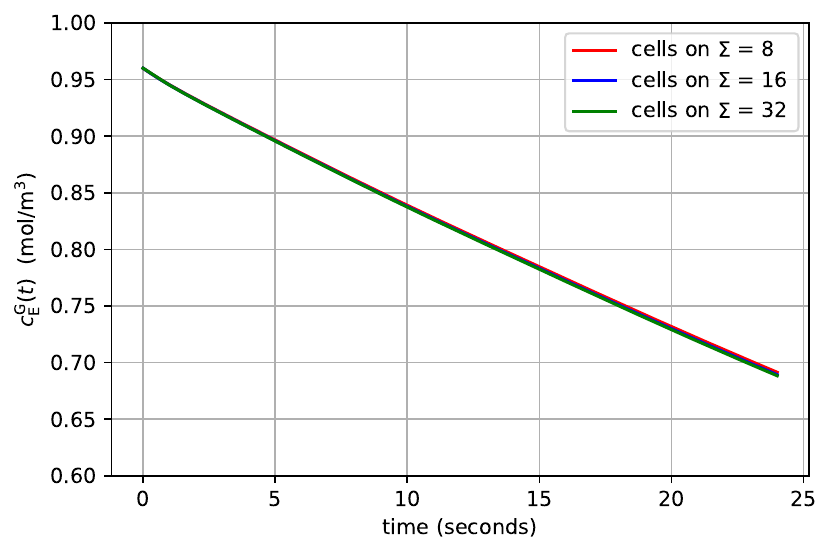}
\caption{Concentration at the interface $\Sigma$ with time.}
\end{subfigure}
\hfill
\begin{subfigure}[htbp]{0.45\linewidth}
\includegraphics[width=\linewidth]{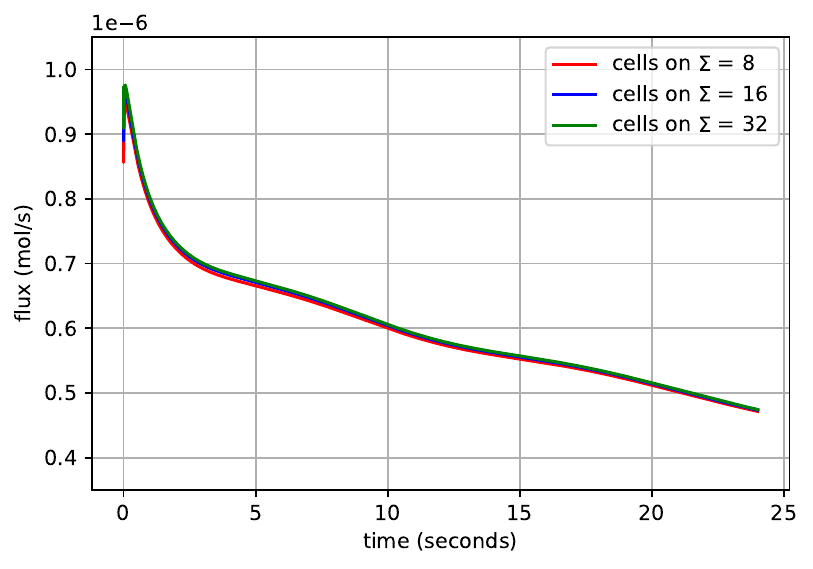}
\caption{Diffusive flux out of the interface $\Sigma$ with time.}
\label{subfig:diffFluxConvergence}
\end{subfigure}\caption{Mesh convergence study on the interface patch.}
\label{fig:convergenceStudy}
\end{figure}

\begin{figure}
    \begin{subfigure}[htbp]{0.45\linewidth}
    \includegraphics[width=\linewidth]{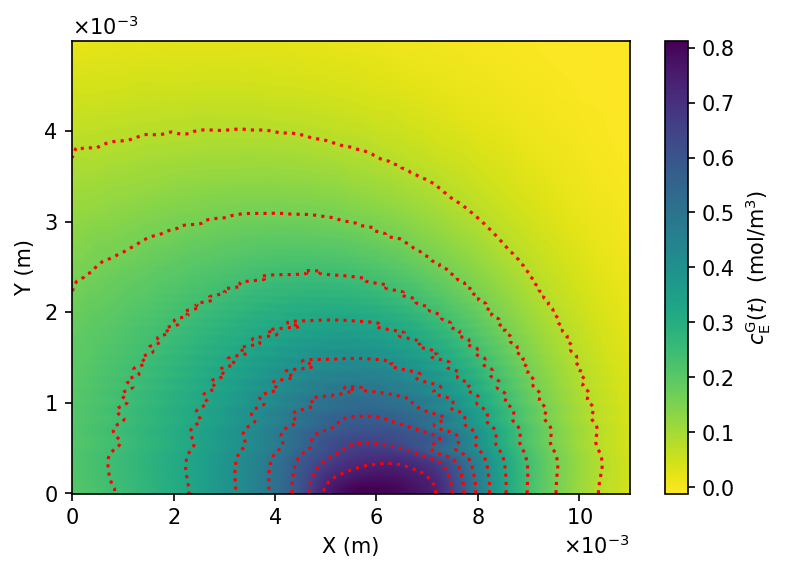}
    \caption{Experimental concentration at time $t=4\,{\rm s}$.}
    \label{subfig:contour_time4s}
    \end{subfigure}
    \hfill
    \begin{subfigure}[htbp]{0.45\linewidth}
    \includegraphics[width=\linewidth]{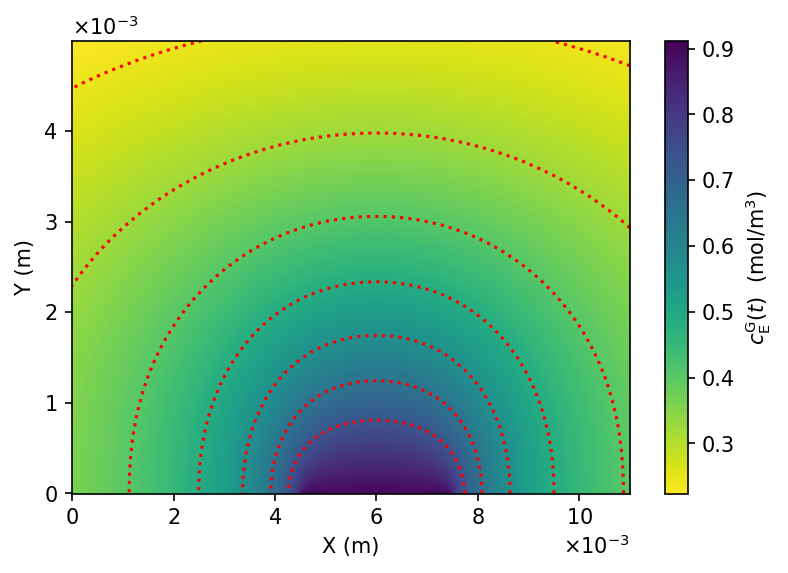}
    \caption{Numerical concentration at time $t=4\,{\rm s}$.}
\end{subfigure}
\begin{subfigure}[htbp]{0.45\linewidth}
    \includegraphics[width=\linewidth]{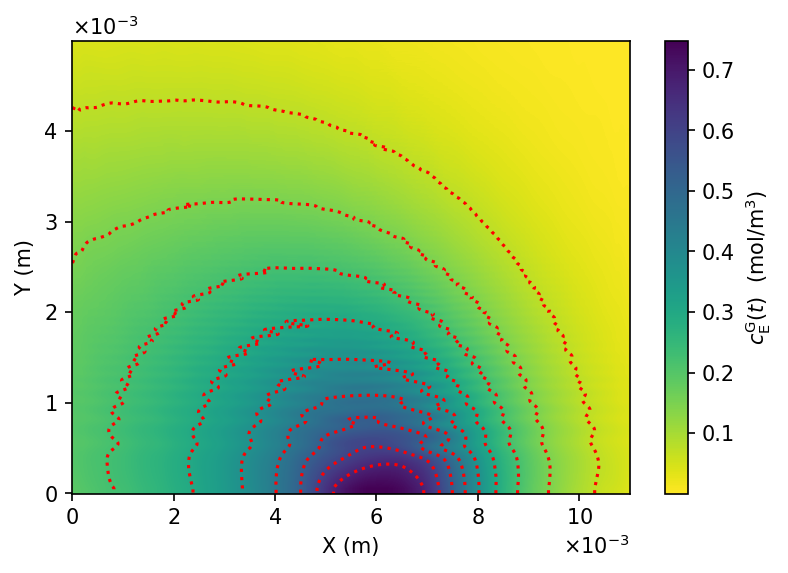}
    \caption{Experimental concentration at time $t=14\,{\rm s}$.}
    \end{subfigure}
    \hfill
    \begin{subfigure}[htbp]{0.45\linewidth}
    \includegraphics[width=\linewidth]{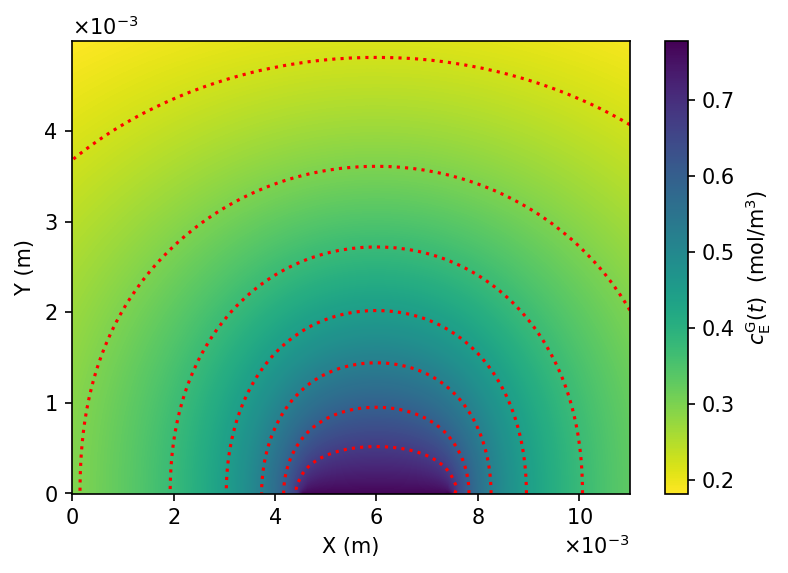}
    \caption{Numerical concentration at time $t=14\,{\rm s}$.}
\end{subfigure}\label{subfig:contour_time14s}
\begin{subfigure}[b]{0.45\linewidth}
        \centering
        \includegraphics[width=\linewidth]{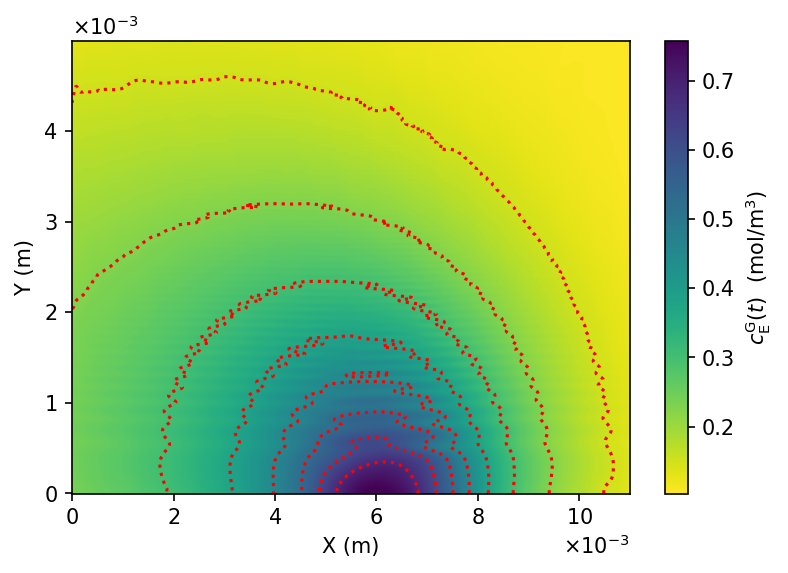}
        \caption{Experimental concentration at time $t=20\,{\rm s}$.}
    \end{subfigure}
    \hfill
    \begin{subfigure}[b]{0.45\linewidth}
        \includegraphics[width=\linewidth]{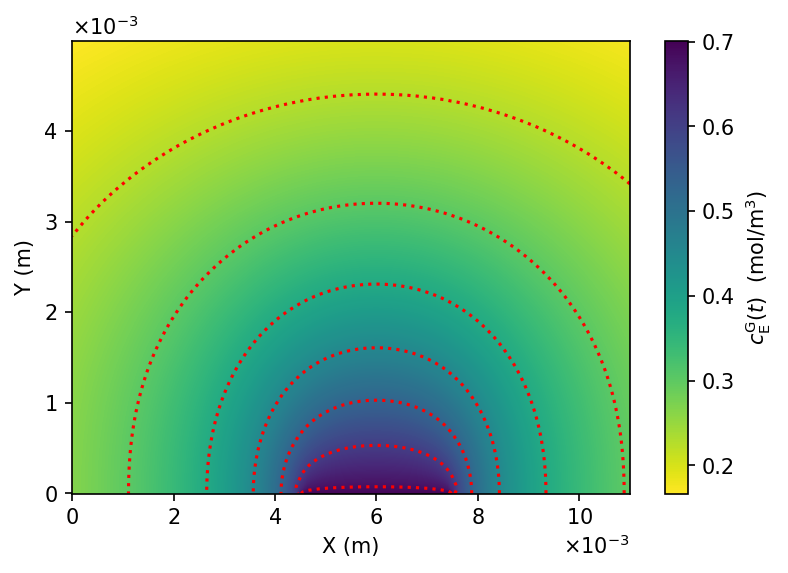}
        \caption{Numerical concentration at time $t=20\,{\rm s}$.}
\end{subfigure}\caption{Qualitative comparison of experimentally measured and simulated vapor field with their iso-contours at different times.}
\label{fig:contour_plots}
\end{figure}

\begin{figure}[ht]
    \centering
    \includegraphics[width=9cm]{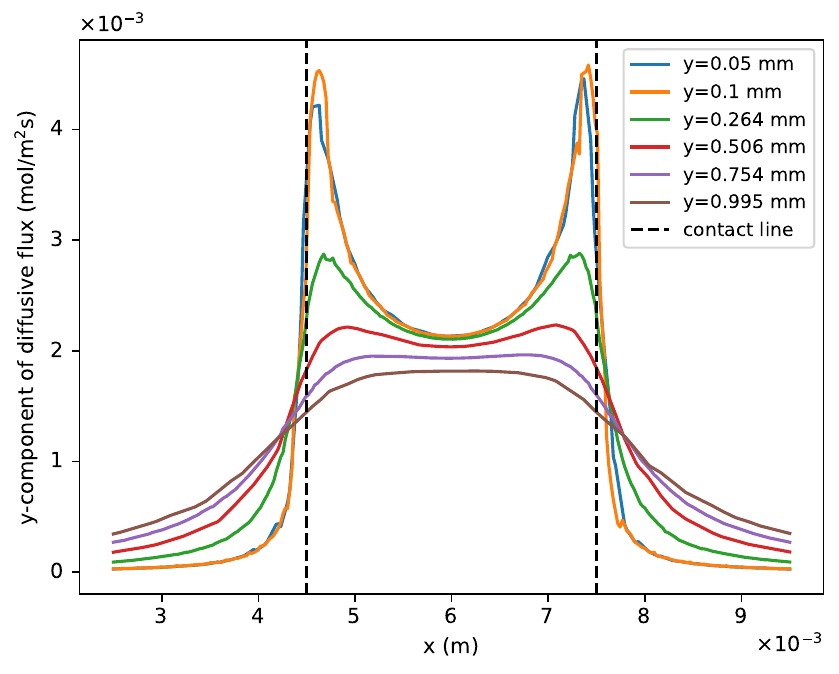}
    \caption{The diffusive flux in $y$ direction is evaluated at different heights above and close to the interface. The wetting edges are identified through the vertical dashed lines where a jump in flux is observed. }
    \label{fig:diffFluxinYdir}
\end{figure}

\begin{figure}[ht]
    \centering
    \includegraphics[width=9cm]{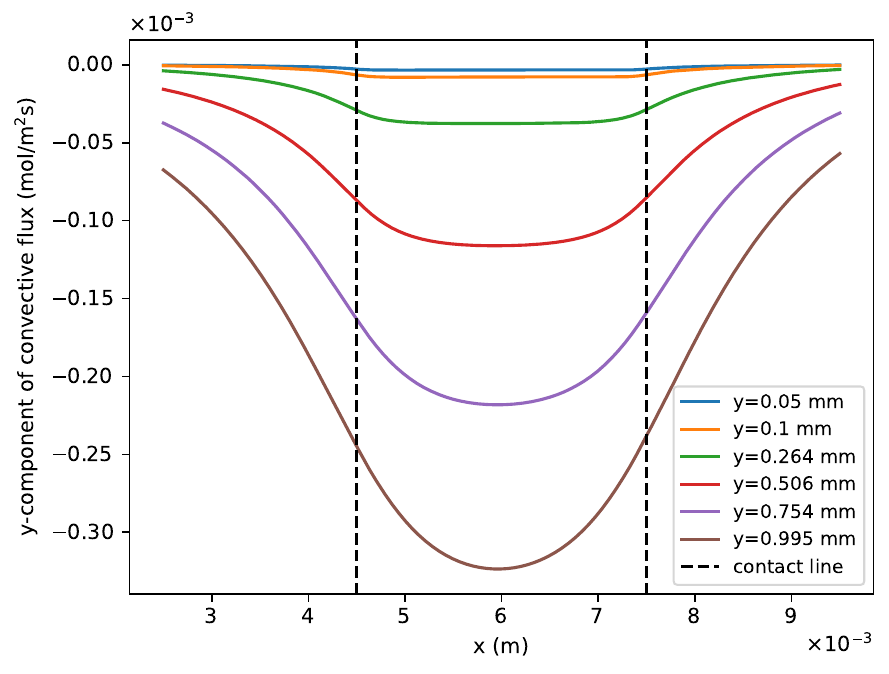}
    \caption{The convective flux in $y$ direction is evaluated at different heights above and close to the interface. The negative values indicate the downward convective flow due to buoyancy. }
    \label{fig:convFluxinYdir}
\end{figure}

\begin{figure}[ht]
    \centering
    \begin{subfigure}[b]{0.45\textwidth}
         \centering
         \includegraphics[width=\textwidth]{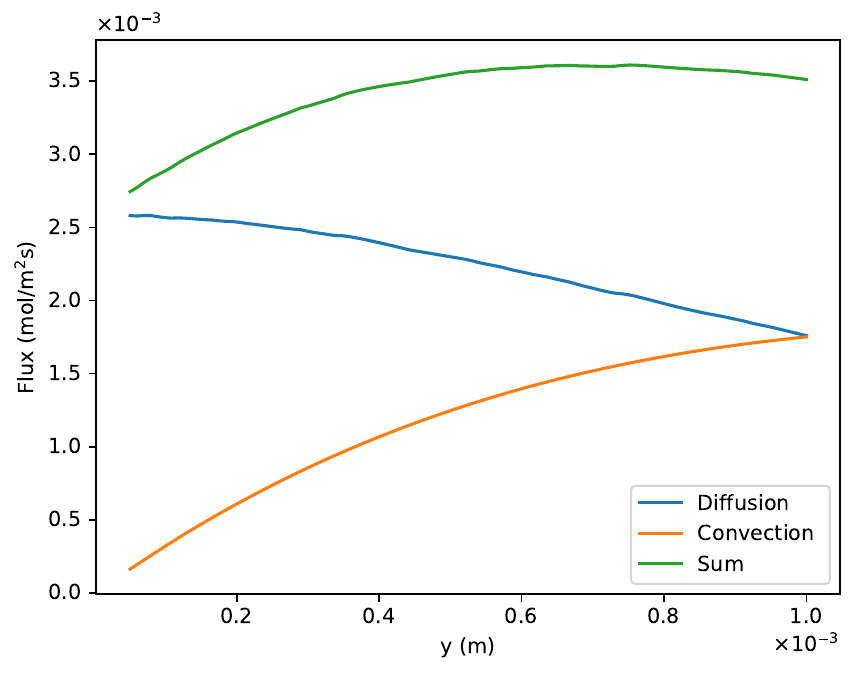}
         \caption{Fluxes in the direction perpendicular to the interface up to $1\;{\rm mm}$ above the interface.}
         \label{fig:lateral_1mm}
    \end{subfigure}
\begin{subfigure}[b]{0.45\textwidth}
         \centering
         \includegraphics[width=\textwidth]{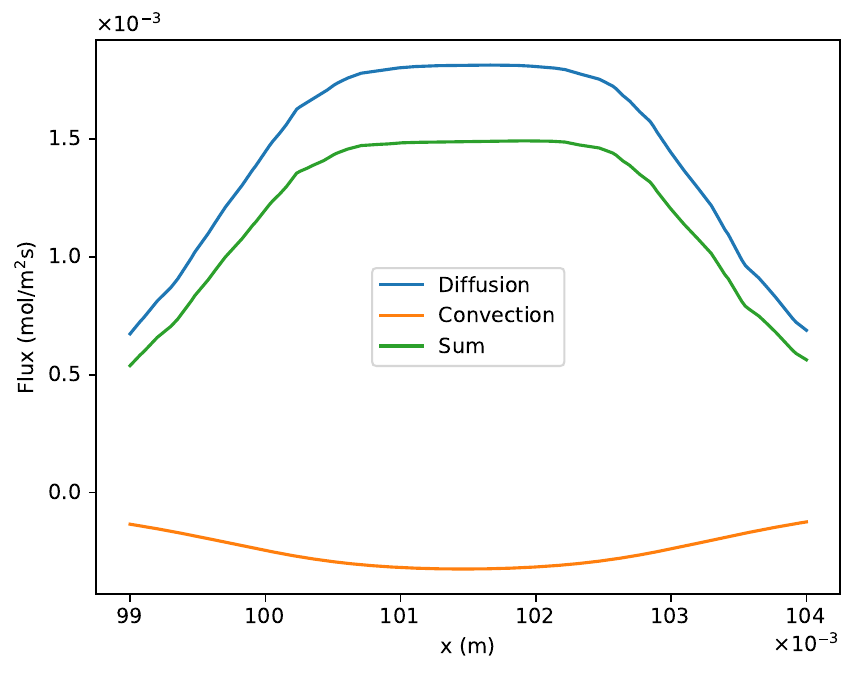}
         \caption{Fluxes in the direction parallel to the interface at height of $1\;{\rm mm}$ from the interface}
         \label{fig:vertical_1mm}
    \end{subfigure}
    \caption{ The net total, diffusive and convective fluxes perpendicular to the interface at two distinct points at a length of $1\;{\rm mm}$ on either side of the contact lines and parallel to the interface at a vertical height of $1\;\text{mm}$. }
    \label{fig:1mmFluxes}
\end{figure}

\begin{figure}[ht]
    \centering
     \begin{subfigure}[b]{0.45\textwidth}
         \centering
         \includegraphics[width=\textwidth]{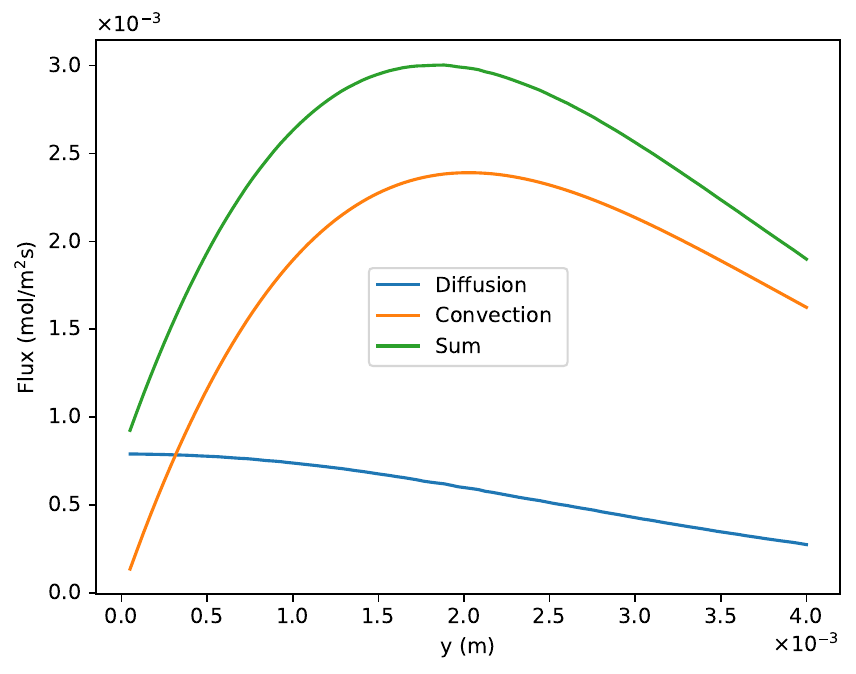}
         \caption{Fluxes in the direction perpendicular to the interface up to $4\;{\rm mm}$ above the interface.}
         \label{fig:lateral_4mm}
     \end{subfigure}
\begin{subfigure}[b]{0.45\textwidth}
         \centering
         \includegraphics[width=\textwidth]{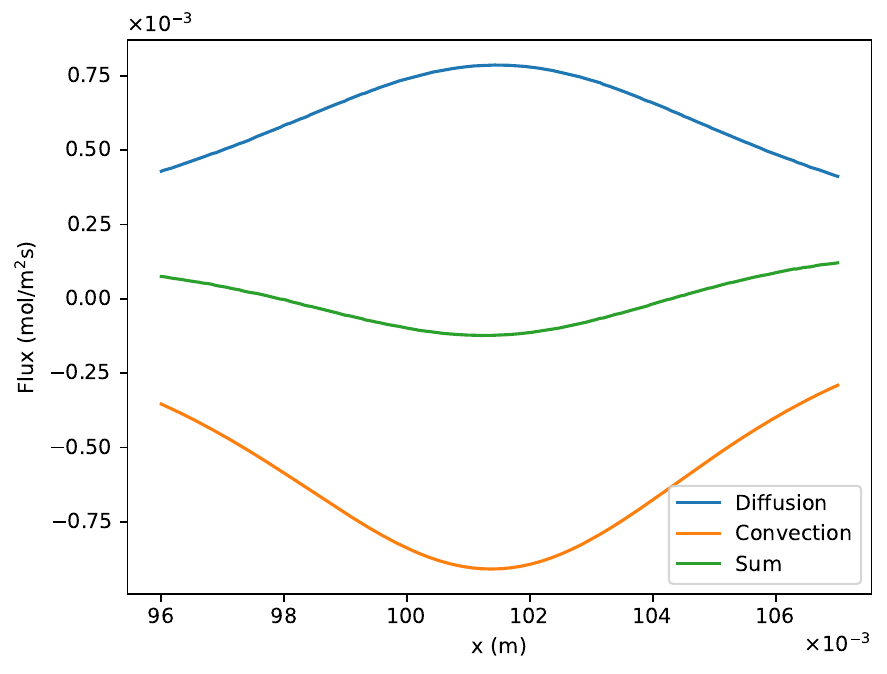}
         \caption{Fluxes in the direction parallel to the interface at height of $4\;{\rm mm}$ from the interface.}
         \label{fig:vertical_4mm}
     \end{subfigure}
    \caption{  The net total, diffusive and convective fluxes perpendicular to the interface at two distinct points at a length of $4\;{\rm mm}$ on either side of the contact lines and parallel to the interface at a vertical height of $4\;\text{mm}$.}
    \label{fig:4mmFluxes}
\end{figure}

The simulation results are compared with mean values of concentration calculated from the experimentally measured data which have an interval of $0.4\;{\rm s}$ between them. The concentration values at each of the pixels are averaged over a length of $1\,{\rm mm}$ along the interface, on either side of its symmetry axis and just above the interface. The initial concentration for the simulation, $\ceth^{\rm G}(0)$ is chosen as this calculated mean value. Figure \ref{fig:conc_0.960C} shows the evolution of concentration at the interface and the value of the calculated modified Henry coefficient, $H_{\rm mod}$. The value of $H_{\rm mod}$ is found out by specifying the value for $\ceth^{\rm G}(0)$ in \eqref{eq:concGasPhase} and is thus used as a fitting parameter. The initial value of concentration, $\ceth^{\rm G}(0) = 0.960\;{\rm mol}/{\rm m}^3$ also implies that about 9 weight $\%$ of ethanol moles are lost initially and that is directly related to the pre-factor value of 0.7 seen in the first equation of \eqref{eq:molesEandEG} in the Appendix \ref{append:reducedConcCalc}. The corresponding diffusive flux from simulations, calculated using Fick's law on the discretised interface is shown in Figure \ref{fig:flux_0.960C}. These values are compared against the diffusive flux evaluated from the experimental data at all the available time intervals. There is an initial fast relaxation occurring in the simulation on the initial vapor field lasting for less than 0.1 second. A closer look is provided in the inset in Figure \ref{fig:fluxEvolution} on this initial behaviour of the flux values. As the experimental concentration field which is used as the initial field in the simulation is subject to measurement errors and also does not extend to the full computational domain, such a fast initial transient is to be expected. This converts the initial 'unphysical' fields into consistent concentration fields by leveling off deviation from local equilibrium. \\
\noindent
The results in Figures \ref{fig:concEvolution} and \ref{fig:fluxEvolution} help to gain insight into the evolution of concentration and the diffusive flux at the film surface when the initial concentration at the interface is varied. This is to understand the impact of variations on the initial condition ($\ceth^{\rm G}(0)$) to account for the uncertainties in the first experimental measurement. Of particular interest is the behavior of the flux values in the initial few time steps, as seen for the curves with $\ceth^{\rm G}(0) = 0.950$ and $\ceth^{\rm G}(0) = 0.980\;{\rm mol}/{\rm m}^3$. The influence of respective $H_{\rm mod}$ are dominant in the initial time steps as well. Even though there are initial differences, towards the end of the simulation time, the concentrations at the interface, as well as the flux values, tend towards each other. 
\newpage
Figure \ref{fig:convergenceStudy} shows a convergence study on the simulation started with $\ceth^{\rm G}(0) = 0.960\;{\rm mol}/{\rm m}^3$. The number of cells discretising the interface is doubled and the concentration and flux is found to be mesh independent. The initial non-physical flux evolution is decreased by a small amount as seen by the length of lines very close to time zero in Figure \ref{subfig:diffFluxConvergence}.
Using the field data from the simulation with the 32 cells on the interface, a qualitative comparison between the experimental and simulation vapor fields is provided in Figure \ref{fig:contour_plots}. Three different time instances are shown. In the experimental vapor field plots, the asymmetry can be seen where the contours are pushed to the left side. This is attributed to a free stream current in the experimental setup and disturbances in the atmosphere due to the movement of the robotic arm which is used to deposit the substrate in the measurement zone. In the simulations, there is a clean interface and no external disturbances apart from smoothing of the initial experimental vapor field input as the starting point for the simulations. So with time, there is a symmetry in the vapor field simulations. All the iso-contours in Figure \ref{fig:contour_plots} are plotted for values from $0.1...0.9$ times the maximum value in the experimental field at the respective times. 
\newline
A verification is done for the flux calculated using Fick's law in the simulations. The procedure used to evaluate the flux from the experimental concentration field is applied on the concentration field from the simulations. Since there is no data available on the convection present in the experiments, we can only evaluate the diffusive flux in a small region around the interface. The flux is evaluated from the field data by integrating along the length of two vertical lines $1\;{\rm mm}$ on either side of the contact line and a horizontal line at same height of $1\rm mm$ encompassing the interface. So in this region the diffusive flux evaluated is exactly same as from Fick's law on the discretised interface. 
\begin{figure}[htbp]
    \centering
    \includegraphics[width=9cm]{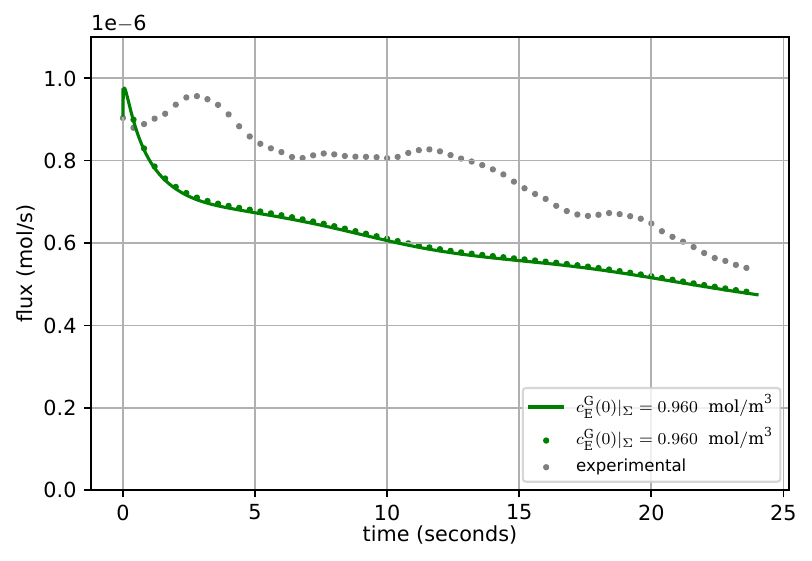}
    \caption{Verification of the diffusive flux from the simulations calculated using Ficks's law.}  \label{fig:diffFluxValidation}
\end{figure}
The result shown in Figure \ref{fig:diffFluxValidation} is further used to provide deeper insights into the physics at the wetting edges of the printed drop and of the vapor atmosphere. Figure \ref{fig:diffFluxinYdir} shows the diffusion flux at different heights above the interface. Wetting edges dispel more diffusive flux into the atmosphere close to the interface. The convective flux is zero in the immediate vicinity of the interface and constitutes less than $10\,\%$ of the diffusive flux but with increasing distance from the interface convection starts to grow as seen in Figure \ref{fig:convFluxinYdir}. The convective flux values are negative owing to the downward motion of the vapor towards the interface, i.e., in the negative $y$ direction. Owing to the structure of the flow inside the simulation domain, the contribution of convection is more in the center of the film. 
The interaction between the convective and diffusive fluxes in the vapor field above the interface can be systematically studied using the rich simulation data. The field data can be divided up into consecutive regions of rectangles. The net fluxes in directions parallel and perpendicular to the interface along these lines which make up the rectangles at different locations in the domain reveal the strength of interactions between the different fluxes. The rectangles fully encompass the interface. Figures \ref{fig:lateral_1mm} and \ref{fig:vertical_1mm} depict the different lateral and vertical direction flux components at a vertical height of $1\;\text{mm}$ above and from the center of the interface. Diffusion dominates in both directions in this vicinity. In Figure \ref{fig:lateral_1mm}, we see the point of crossover of the diffusive and convective fluxes occur at a height of about 1mm from the interface. The sum of fluxes in the region takes the shape of the diffusive flux.
Moving further away at a height of $4\;\text{mm}$ from the interface, Figure \ref{fig:4mmFluxes} shows that the sum of fluxes takes the shape of the dominating convective flux and the crossover regions of the two fluxes can be further identified in Figure \ref{fig:lateral_4mm}. 

\newpage
\clearpage
\section{Discussion}
\label{discussion}

In our study, we have presented a combined method for measurement of vapor pressure distributions and simulation-based reconstruction of vapor flows, based on the boundary conditions obtained from experimental data. In this way, we could distinguish the diffusive vapor flow from the convective one which is driven by the buoyancy effects of ethanol vapor which has a mass density that is 50~\% larger than that of air. 
We could also identify some points which call for closer discussion and interpretation.
\begin{itemize}
\item {\em Calculation of the vapor flux} in the gas atmosphere, based on vapor concentration measurement, was possible with our setup. This is important because we can check the mass balances of the evaporation process, and predict the remaining concentrations of the volatile component in a liquid. We also found that it is essential to distinguish between diffusive and convective fluxes. This cannot be done by vapor pressure measurement alone, but requires the reconstruction of the flow velocities by numerical means. Evaporation of a relatively dense vapor eventually induces convection by buoyancy effects, which changes the conditions of diffusion. This, in turn, accelerates the evaporation fluxes.
\item {\em Evaporation from mixtures}: By measuring the vapor pressure close to the liquid gas interface, it is possible to determine the local concentration of the volatile liquid component in the mixture. This yields important information on the local composition of the liquid film, and, via a possible Marangoni effect, on the internal liquid transport. We did not include this in our simulations, but it has become apparent how this could be done in the future.
\item {\em Edge enhancement of evaporation} has been observed in experiment as well as in simulation, and the local vapor concentration profile in the vicinity of the edges of the liquid film could be resolved with good agreement. This is a clear strength of our 2D approach to the topic. It is appealing to map these profiles with more details models in the future.
\item {\em Reducing evaporation to 2D}, however, raises a principal issue regarding the relation of diffusion and convection. If the vapor transport was purely diffusive, no physically meaningful steady-state of vapor transport could be attained. Vapor pressure profile and flow density would essentially depend on the boundary conditions in far distance, and this cannot be related to observation, of course. The experiment, even though tailored to 2D close to the evaporating liquid layer, finally becomes 3D in large but finite distance, and thus achieves an asymptotic dilution of the vapor. However, the diffusion law in 2D, namely $-\Delta\ceth^{\rm G} \sim \delta(\boldsymbol{r})$, predicts that vapor concentration $\ceth^{\rm G}(r)$ must decrease like $\ceth^{G}(r)\sim \log\,(r^{-1})$ with distance $r$ from the evaporation source at the origin, and not with $r^{-1}$ as in 3D. Consequently, the 2D simulation cannot be matched with an asymptotically vanishing vapor concentration at infinity without spoiling the boundary conditions. In order to keep up vapor flux from the source to an infinite sink in the 2D simulation, we required the instant $t_{\rm conv}$ of transition to convective vapor transport to occur within the simulation time interval $T$. This creates the necessary asymptotic vapor dilution at infinity. More precisely, it was required that $t_{\rm conv}^2 < L_{\Omega}/\Ds$, where $L_{\Omega}$ is the size of the simulation domain $\Omega$. This cannot be guaranteed in advance, but depends on the dynamics that creates the convective flow from the buoyancy forces of the evaporating material, i.e.\ diffusion timescale $(L_{\Omega}/\Ds)^{1/2}$ in the simulated domain had to be larger than the timescale of developing convection.
\item {\em Evaporation from a two-component liquid film} requires, in principle, also a modeling and simulation of the liquid film dynamics itself. Evaporation would lead to the formation of concentration gradients in the liquid film, which, in turn, could drive Marangoni flows and liquid-film-instabilities. We have blanked out this feature, and applied a strongly simplified model for the evaporating liquid film. It has become quite apparent that a more dynamic model of liquid film behavior could offer principal improvements. 
\item {\em Vapor pressure reduction}: For the majority of liquid mixtures, vapor pressure reduction of the volatile component is a feature which is not sufficiently described by Raoult's law. Rather, the correction by Henry's law appears substantial. However, this will only be successful for a limited concentration range. In our case, we could restrict to small ethanol concentrations $\xethL < 0.3$ where this simple correction was feasible. If one is interested in a larger range of concentrations, a more detailed model of the vapor pressures appears to be advised. 
\item{\em Evaporation resistivity} is another mechanism which decreases the vapor pressure in the gas atmosphere. We could uniquely distinguish it from a possible effect of vapor pressure reduction by measuring the total amount of evaporated material from measurement as well as from simulation, and cross-com\-pari\-son of the mass balances in the liquid film. The observation of an evaporation resistance implies that there is some contamination or adsorption layer of impermeable material on the liquid-gas interface. The measured vapor pressure (and thus the chemical potential) of the volatile component close to the interface at some given evaporation flow density (i.e. apart from thermal equilibrium) can be used to determine the strength of the evaporation resistivity, provided that the chemical potential of the component in the liquid film is known as well.  
\item {\em Particular physico-chemical aspects of surfaces}: The question arises on the nature of the observed impurities responsible for evaporation resistivity. There are numerous possible reasons, such as silicone or tenside residuals in the inkjet print system. However, it seems likely that in our particular experiment the aluminium substrate of the liquid film was the source of contamination. This is because both contact area and time with the ethanol mixture were very large. Moreover, it is known that aluminium metal is not strictly inert with respect to ethanol, and that aluminium ethoxide may form in presence of small traces of certain ions as, e.g., iodine \cite{Wilhoit1962}. This soluble salt could gather at the liquid-air interface and accumulate to a more or less dense layer or skin of insoluble aluminium hydroxide. This could tentatively explain the observed evaporation resistance. 
\end{itemize}
 
\section{Conclusion}
\label{conclusion}

Understanding the physics of the gas atmosphere over an evaporating liquid layer is a key feature of contemporary industrial production, and in particular in large-scale printing and solvent-based coating technologies. The monitoring of vapor concentrations, and the reconstruction of the flux profiles could foster important progress towards a more efficient and resource-saving technology.

For economical and ecological reasons, the development favors the use of water-compatible solvents, and strives for the replacement of hydrocarbon and aromatic materials. One may be tempted to summarize that solvent-based printing and coating processes tend to develop towards more polar materials, replacing alcane derivatives and aromatic process liquids by water, ethanol, isopropanol. This transition provides many challenges for the engineer, however, there also arise new options for gas-phase process control, which we would like to work out. One primary observation is that polar liquids and their vapors have a molar refractivity which is superior to, e.g., alcanes or toluene by a factor of three to four, and the optical contrast versus air is increased to an even more sensible level. This forwards laser interferometric imaging methods to a technological position that it did not have before. 

Another important application is the monitoring of water vapor in a solvent-based printing or coating process. Molar optical refraction of water ($3.78\,{\rm cm}^3/{\rm mol}$) is slightly smaller than that of air ($4.4\,{\rm cm}^3/{\rm mol}$), and its optical contrast in the interferometer is only 8 \% of that of ethanol under equal conditions. Note that the phase shift of water vapor in air obtained from eq. (\ref{Ev_theo_11}) has the opposite sign. Nevertheless, in\-ter\-fero\-met\-ric water vapor imaging appears feasible with an in\-ter\-fero\-me\-ter of moderately larger base length, or, with the setup used here with reduced resolution of vapor concentration. Industrial drying processes of water-based inks frequently apply heat to the surface, and the specific heat of evaporation is considerable larger than for any other liquid. Although the molar refraction of gases and vapors is practically independent of temperature, the tracking of gas flows by simulation is more challenging by the feature that temperature and energy flows must be included as well. 

Moreover, it was not by accident that we were interested in a 2D projection of the evaporation process, as this is the natural constellation along a continuous web-fed manufacturing line. Imaging and processing directions should, of course, coincide. The laser in\-ter\-fero\-me\-ter then perceives a more or less steady-state situation over the possibly rapidly moving web, with all the diffusion-limited and advection-driven vapor (and heat) transport phenomena. Our somewhat artificial restriction to evaporation from rectangular wetting stripes, with a cumbersome correction strategy for the finite-length effects, becomes obsolete there. It is also possible to employ our system in a non-isothermal environment. However, the primary quantity of imaging is not the partial vapor pressure but vapor density, which is dependent on both local vapor pressure and temperature.

When designing our interferometric setup, we have taken particular care to use optical and electronic components with good commercial availability, and which also appear adequate to industrial standards. The only particular feature is the robot platform, but this is due to our specific lab restrictions. One could well imagine this with quite different machine concepts.

\paragraph{Acknowledgments:} We acknowledge the financial support by the German Research Foundation (DFG) within the Collaborative Research Centre 1194 (Project-ID 265191195).

\bibliographystyle{ieeetr}

\newpage
\clearpage
\section{Appendix}
\label{sec:appendix}
\subsection{Appendix 1}
\label{app:constBuoyancyTerm}
The mass density of a mixture of air with average molar mass of $M_{\rm air} = 0.028\,{\rm kg}/{\rm mol}$ and ethanol vapor with $M_{\rm E} = 0.046\,{\rm kg}/{\rm mol}$ is a function of the sums of their partial pressures, weighted with their molar masses. This is because their molar volume, according to the ideal gas equation, is the same at given total atmospheric pressure $p_{\rm tot} = 100,100\,{\rm Pa}$ and temperature $T = 303\,{\rm K}$. If the partial pressure of ethanol vapor is $p_{\rm E}$, the partial pressure of air must be $p_{\rm tot} - p_{\rm E}$. From this, it follows:
\begin{equation*}
    \begin{aligned}
        \rho(c_{\rm E}) =& ~ \frac{1}{RT}\,\left(M_{\rm air} (p_{\rm tot}-p_{\rm E}) \,+\, M_{E} p_{\rm E}\right)\\[2mm]
        =& ~ \frac{1}{RT}\,\left(M_{\rm air} p_{\rm tot} \,+\, \left[M_{\rm E}\,-\,M_{\rm air}\right] p_{\rm E}\right)\\[2mm]
        =& ~ \rho_{0}\,+\,\frac{M_{\rm E}\,-\,M_{\rm air}}{RT}\,p_{\rm E} \end{aligned}
\end{equation*}
with $\rho_{0} = M_{\rm air} p_{\rm tot} / RT$ as the density of pure air at the respective temperature. We know the upper bound of the saturated ethanol vapor concentration in equilibrium with pure liquid ethanol which is $0 \le c_{\rm E} \le 3.53\,{\rm mol}/{\rm m}^3$ having a partial pressure in $0 \le p_{\rm E} \le 8,895\,{\rm Pa}$ at $303\,{\rm K}$.\\
\noindent
According to the ideal gas equation, applied to the (very diluted) ethanol vapor, the molar concentration $({\rm mol}/{\rm m}^3)$ is
\begin{equation*}
    c_{\rm E}\;=\;\frac{p_{\rm E}}{RT} .
\end{equation*}
This identity is independent of any evaporation features and other properties of the liquid phase coexisting with the vapor. Thus one finds that
\begin{equation*}
    \rho(c_{\rm E}) = \rho_{0} + \left(M_{\rm E} - M_{\rm air}\right) c_{\rm E} .
\end{equation*}
\subsection{Appendix 2}
\label{append:reducedConcCalc}
To calculate the time dependent concentration of ethanol on the interface in the gas phase, we start with re-writing equation \eqref{eq:timeDrichlet} in terms of moles of ethanol in the liquid phase varying with time $\moleE(t)$. So \eqref{eq:timeDrichlet} is rewritten as,
\begin{equation}
    \moleE(t)=\moleE(0)-\frac{1}{h_{\rm film}} \int \displaylimits_0^t \frac{1}{| \Sigma |} \int \displaylimits_{\Sigma} \Ds \nabla c_{\rm E}\left(\boldsymbol{r},t^{\prime}\right) \cdot \intN \mathrm{d} o \mathrm{d} t^{\prime} .
        \label{eq:molesInTime}
\end{equation}
To calculate the initial number of moles of the components in the mixture specifically $\moleE(0)$ , we assume that the components add up linearly to form the liquid mixture i.e.,
\begin{equation}
    (\massfracE(0), \VliqE(0)) + (\massfracEG(0), \VliqEG(0))  = (m_{\rm mix}(0), V_{\rm mix}(0))
\end{equation}
where $\massfracE(0)$ and $\massfracEG(0)$ are mass fractions of ethanol and ethylene glycol at time zero respectively and $\VliqE$ and $\VliqEG$ are the volume fractions of ethanol and ethylene glycol respectively. The total mass in the liquid film is then $m_{\rm mix}(0) = \massfracE(0) + \massfracEG(0)$ and $V_{\rm mix}(0) =  \VliqE(0) + \VliqEG(0)$ is the total volume of the liquid  mixture. Considering the values of the mass fractions of ethanol and ethylene glycol initially in the mixture, the ratio of the respective volumes in the liquid gives 
\begin{equation*}
    \frac{\VliqE(0)}{\VliqEG(0)} = \frac{0.3}{0.7} \frac{\rhopureEG}{\rhopureE} \coloneqq \lambda_0 .
\end{equation*}
With the above definition, the volumes and correspondingly the mass of ethanol and ethylene glycol in the printed liquid film can be estimated as,
\begin{equation}
    \begin{aligned}
        \VliqE(0) = \frac{\lambda_0}{1+\lambda_0} V_{\rm mix}(0) , \\
        \massfracE(0) = \rhopureE \VliqE = \frac{0.70 \rhopureE \lambda_0}{1+\lambda_0} V_{\rm mix}(0)
    \end{aligned}
\end{equation}
for ethanol in the liquid. A pre-factor value of 0.70 in the above equation indicates that certain amount of ethanol out of the 30 weight percent ethanol in the mixture initially printed are lost before the beginning of the measurement and correspondingly the simulation. This corresponds to 9 weight $\%$ ethanol moles lost at beginning of measurement which is an approximation. Similarly, the mass fraction of ethylene glycol in the printed liquid film is 
\begin{equation}
    \begin{aligned}
        \VliqEG(0) = \frac{1}{1+\lambda_0} V_{\rm mix}(0) , \\
        \massfracEG(0) = \rhopureEG \VliqEG = \frac{\rhopureEG}{1+\lambda_0} V_{\rm mix}(0) .
    \end{aligned}
\end{equation}
The number of moles of ethanol and ethylene glycol thus present can be calculated as,
\begin{equation}
    \begin{aligned}
        \moleE(0) = \frac{0.70 \rhopureE \lambda_0}{(1+\lambda_0)\massfracE} V_{\rm mix}(0), \\
        \moleEG(0) = \frac{\rhopureEG }{(1+\lambda_0)\massfracEG} V_{\rm mix}(0) .
    \end{aligned}
    \label{eq:molesEandEG}
\end{equation}
Together with equations \eqref{eq:molesInTime} and \eqref{eq:molesEandEG}, the initial concentration at $t=0$ and concentration at the interface with time $t$ in the gas or vapor phase with time can be calculated as, 
\begin{equation}
    \begin{aligned}
        \ceth^G(t) = \frac{1}{RT} \psat \xethG(t), \\
        \ceth^G(t) = \frac{\psat}{RT} H_{\rm mod} \molefracE(t), 
    \end{aligned}
\end{equation}
where $\molefracE(t) = \moleE(t) / (\moleE(t) + \moleEG(t))$ is the mole fraction of ethanol in the liquid phase with time, $\xethG(t)$ is the mole fraction of ethanol in the gas phase,  $\psat$ is the saturation pressure of pure ethanol, $R$ is the universal gas constant, $T$ is the temperature and $H_{\rm mod}$ is the modified Henry coefficient which is a product of evaporation resistivity factor $\alpha_{\rm E}$ and Henry coefficient $H$. Equation \eqref{eq:concGasPhase} along with \eqref{eq:molesInTime} is updated at the end of every time step while \eqref{eq:molesEandEG} is used at the beginning of the simulation at time $t=0$. 
 
\end{document}